\documentclass[twocolumn,tighten,times]{aastex701}
\submitjournal{Nature Communications}

\begin{document}

\title{JWST interferometric imaging reveals the dusty disk obscuring the supermassive black hole of the Circinus galaxy}

\author[0000-0001-5357-6538]{Enrique Lopez-Rodriguez}
\affiliation{Department of Physics \& Astronomy, University of South Carolina, Columbia, SC 29208, USA}
\affiliation{Kavli Institute for Particle Astrophysics \& Cosmology (KIPAC), Stanford University, Stanford, CA 94305, USA}
\email[show]{elopezrodriguez@sc.edu}

\author[0000-0002-9723-0421]{Joel Sanchez-Bermudez}
\affiliation{Instituto de Astronom\'ia, Universidad Nacional Aut\'onoma de M\'exico, Apdo. Postal 70264, Ciudad de Mexico 04510, Mexico}
\email[]{joelsb@astro.unam.mx}  

\author[0000-0002-2356-8358]{Omaira Gonz\'alez-Mart\'in}
\affiliation{Instituto de Radioastrononom\'ia y Astrof\'isica (IRyA), Universidad Nacional Aut\'onoma de M\'exico, Antigua Carretera a P\'atzcuaro \#8701, 1087 ExHda. San Jos\'e de la Huerta, Morelia, Michoac\'an, M\'exico C.P. 1088 58089}
\email[]{o.gonzalez@irya.unam.mx}

\author[0000-0002-7052-6900]{Robert Nikutta}
\affiliation{NSF NOIRLab, 950 North Cherry Avenue, Tucson, AZ 85719, USA}
\email[]{robert@nikutta.org}

\author[0000-0003-0778-0321]{Ryan M. Lau}
\affiliation{NSF NOIRLab, 950 North Cherry Avenue, Tucson, AZ 85719, USA}
\email[]{ryan.lau@noirlab.edu}

\author[0000-0002-1536-7193]{Deepashri Thatte}
\affiliation{Space Telescope Science Institute, 3700 San Martin Drive, Baltimore, MD 21218, USA}
\email[]{thatte@stsci.edu}

\author[0000-0002-9627-5281]{Ismael Garc\'ia-Bernete}
\affiliation{Centro de Astrobiolog\'ia (CAB), CSIC-INTA, Camino Bajo del Castillo s/n, E-28692 Villanueva de la Ca\~nada, Madrid, Spain}
\email[]{igbernete@gmail.com}

\author[0000-0001-8627-0404]{Julien H. Girard}
\affiliation{Space Telescope Science Institute, 3700 San Martin Drive, Baltimore, MD 21218, USA}
\email[]{jgirard@stsci.edu}

\author[0000-0001-9315-8437]{Matthew J. Hankins}
\affiliation{Arkansas Tech University, 215 West O Street, Russellville, AR 72801, USA}
\email[]{mhankins1@atu.edu}

\begin{abstract}

The dusty and molecular torus is one of the most elusive structures surrounding supermassive black holes, yet its importance is unequivocal for understanding feedback and accretion mechanisms. The torus and accretion disk feed the inspiraling gas onto the supermassive black hole (SMBH) and launch outflows, fundamentally connecting the SMBH activity to the host galaxy. This scenario situates the torus as the interface between the AGN and its host galaxy with a flow cycle of molecular gas and dust of a few parsecs in size. 
Here, we utilize a novel aperture-masking interferometric mode onboard the JWST, achieving twice the previously possible resolution, and bringing out the fainter features that clearly show the torus being the critical interface for feeding material from galaxy scales into the SMBH. 
We also identify that $<1$\% of the emission arises from an arc structure composed of hot dust entrained in a molecular and ionized outflow. The rest of the emission, $12$\%, is associated with dust heated by the AGN and/or radio-jet at large scales. Combined with continuum data, gas tracers, and torus models, our study shows that most of the dust mass is located in the equatorial axis in the form of a disk feeding the AGN.

\end{abstract}



\section*{\textbf{Introduction}} \label{sec:Intro}

The buildup of the central supermassive black hole (SMBH) mass is a fundamental facet of galaxy growth and evolution and occurs at least in part through active galactic nuclei (AGN) accretion in massive galaxies. On the other hand, AGN feedback via wide-angle winds can quench \citep{Morganti2017} and/or enhance \citep{Cresci2015} star formation in galaxies \citep{Morganti2017}, regulate SMBH accretion \citep{Hopkins2016}, and even shape galaxy morphology \citep{Dubois2016}. The characterization of the central $1-100$ pc around AGN reveals the origins of accretion and feedback mechanisms during galaxy evolution and provides constraints for theoretical models of galaxy formation.

The bulk of dust and molecular mass cospatial with accreting and outflowing material is studied at sub-mm wavelengths using ALMA \citep[Atacama Large Millimeter/submillimeter Array;][]{GB2024}. Further mid-infrared (MIR; $7-12\,\mu$m) interferometric observations are more sensitive to the extended emission above and below the disk. This MIR emission predominantly arises from warm, optically thin dust layers in the torus walls and/or a dusty wind \citep{Honig2019,GR2022} launched by radiation pressure from a magnetohydrodynamical wind generated at sub-pc scales \citep{Emmering1992,ELR2015,ELR2023,Gallimore2024}

The near-IR (NIR; $1-5\,\mu$m) observations are sensitive to the hot dust ($400-1500$ K), which traces the inner edge of the torus and/or the base of the torus walls or dusty winds \citep{Kishimoto2009,Honig2019,Kishimoto2022}. Surprisingly, the physical structures that produce the measured $3-5\,\mu$m excess emission in AGN since first observed in the early '90s still remains unclear \citep{PK1993,Mor2009,AH2011,GB2022}. This observed NIR excess is thought to arise either from hot dusty winds, hot graphite dust in the inner torus, and/or residual starlight from the host galaxy. Unequivocally, any of these scenarios have critical consequences on the accretion and feedback mechanisms on the buildup of the central SMBH. Thus, the identification of the physical structure and mechanism producing the NIR excess will allow us to connect the accretion disk with the reservoir of gas feeding it and the interaction with the host galaxy.

The Circinus galaxy is the best candidate to solve a major issue in AGN physics within the central $10$ pc: wind (outflow) vs. torus (accretion). Circinus hosts the nearest, $4.2\pm0.7$ Mpc \citep[$0.1" = 2$ pc;][]{Tully2009}, type 2 Seyfert galaxy with an intermediate bolometric luminosity of L$_{\rm{bol}} = 10^{43.6}$ erg s$^{-1}$ \citep{Moorwood1996,Oliva1999,Tristram2007}, and shows well-defined inflowing and outflowing dusty and molecular material from the host spiral galaxy and its AGN \citep{Izumi2023}.  It has a kpc-scale radio jet at a PA\,$\sim-64^\circ$ \citep{Elmouttie1998} and a pc-scale radio jet at a PA\,$\sim-84^\circ$ \citep{Izumi2023}.

Circinus has a compact $\sim1.9$ pc diameter central disk and a $\sim5$ pc extended diffuse dusty component perpendicular to the maser disk as observed at $4.7\,\mu$m ($4$ mas resolution) and N-band ($10\,\mu$m; $9$ mas resolution) using observations with the the Multi-aperture Mid-Infrared Spectroscopic Experiment (MATISSE) on the Very Large Telescope Interferometer (VLTI) \citep{Isbell2022,Isbell2023}. A $30\times4$ mas$^{2}$ ($0.6\times0.08$ pc$^{2}$) disk-like structure at a position angle of $\sim46^{\circ}$ East of North dominates the emission at $4.7\,\mu$m MATISSE/VLTI \citep{Isbell2023}. This structure only accounts for $<5$\% of the measured flux within the $0.4"$ aperture measured by the Nasmyth Adaptive Optics System-Coude Near Infrared Camera (NACO) on the VLT in the L-band ($4.8\,\mu$m). However, the origin of the extended emission within $0.1-0.5"$ ($2-10$ pc), still missing from the MATISSE/VLTI observations, and the total emission of directly heated dust in the torus and/or winds remains unknown.

The MIR ($7-12\,\mu$m) emission extends up to $\sim20$ pc on only one side of the ionization cone walls as observed using the VLT spectrometer and imager for the mid-infrared (VISIR) \citep{Stalevski2017}. The $5$ pc-scale dusty emission extension has a wider opening angle and is clumpier than the $20$ pc-scale structure. The N-band observations have been characterized using a series of torus-only and torus+winds models, concluding that the most likely scenario is given by a clumpy disk+hyperboloid component \citep{Stalevski2017}. Under this scenario, the warm ($100-300$ K) dusty extended emission is thought to arise from anisotropic radiation at the location of the maser disk \citep{Greenhill2003}. The $1-20\,\mu$m spectral energy distribution (SED) shows that the N-band observations strongly drive the model fit. However, models are highly degenerate when explaining the hot ($>400$ K) dust component at NIR wavelengths. It urges us to quantify the morphology and properties of the hot dust in the torus and extended structures, and identify the dominant physical structure responsible for the hot and warm dusty extended emission within the central $10$ pc of AGN.


\section*{\textbf{New JWST interferometric observations and images}} \label{sec:OBS}

\begin{figure*}[ht!]
\includegraphics[width=\textwidth]{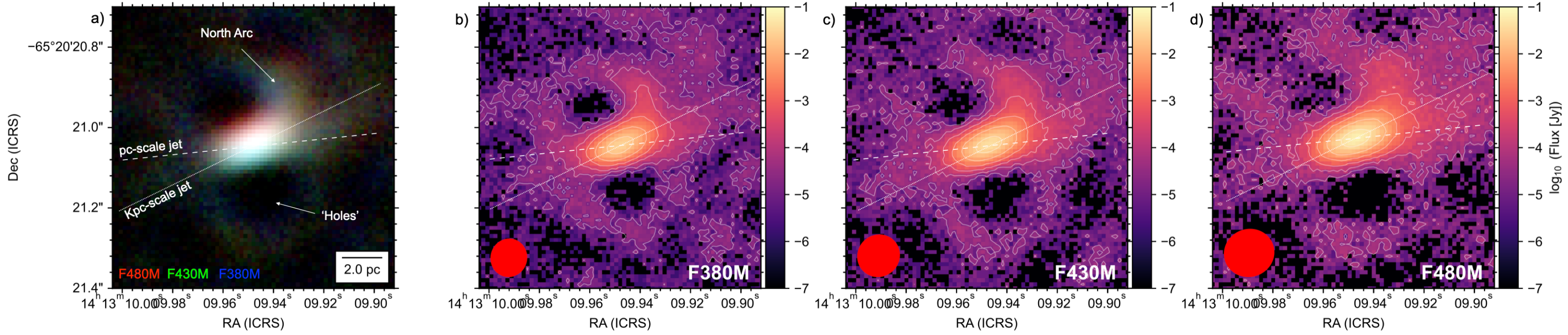}
\caption{\textbf{The dust emission of the central $14\times14$ pc$^2$ of the Circinus galaxy observed with AMI/JWST.}
\textbf{(a)} The RGB image (Red: F480M, Green: F430M, Blue: F380M) of Circinus with the orientations of the kpc-scale \citep[dotted line, PA $= -64^{\circ}$;][]{Elmouttie1998}, pc-scale \citep[dashed line, PA $=-84^{\circ}$;][see Fig. \ref{fig:fig2}]{Izumi2023} radio jets, the `North Arc', and `Holes' features. A 2 pc scale is shown. 
\textbf{(b-d)} The AMI/JWST observations of the continuum dust emission at $3.8\,\mu$m\,(F380M), $4.3\,\mu$m\,(F430M), and $4.8\,\mu$m\,(F480M). The beam sizes (red ellipses) of $93\times88$ mas$^{2}$ ($1.9\times1.8$ pc$^{2}$), $105\times101$ mas$^{2}$ ($2.1\times2.0$ pc$^{2}$), and $123\times116$ mas$^{2}$ ($2.5\times2.3$ pc$^{2}$) are shown in each panel. The contours start at $4\sigma$ and increase in steps of $2^{n}\sigma$ with $n=2,4,6,\dots$. Dec., declination; RA, right ascension.
 \label{fig:fig1}}
\end{figure*}

We observed the Circinus galaxy on July 2024 and March 2025 with the Aperture Masking Interferometry \citep[AMI;][]{AMI2023} mode in JWST's Near Infrared Imager and Slitless Spectrograph \citep[NIRISS;][]{NIRISS2012} at $3.8~\mu$m (F380M), $4.3~\mu$m (F430M), and $4.8\,\mu$m (F480M) (Methods section `Observations'). Both observations ensure a $\sim90^{\circ}$ rotation of the uv-plane to increase its coverage, minimizing image reconstruction artifacts. The $\sim65$ mas NIRISS pixels are Nyquist sampled at $\sim4\,\mu$m in the medium band. Interferometric observables, closure phases and square visibilities \citep{Jennison1959,Readhead1988}, are extracted from the calibrated data and used for image reconstruction. We used SQUEEZE \citep{squeeze2010} for the image reconstruction. Additionally, we performed a bootstrapping analysis on the uv-plane to obtain the significance level of the features in our final images (Methods section `Image reconstruction'). We used the peak emission from the interferogram images to assign the world coordinate system in the reconstructed images (Methods section `WCS correction'). A standard star with known IR fluxes, previously used to perform the flux calibration of the MATISSE/VLTI observations of Circinus \citep{Isbell2022,Isbell2023}, was observed after the Circinus observations and used to perform the flux calibration at each filter (Methods section `Flux calibration'). We estimate that emission lines have small contributions, $<10$\%, within the AMI filters (Methods section `Emission line contribution'). Fig. \ref{fig:fig1} shows the SQUEEZE final reconstructed images of Circinus with angular resolutions ($\lambda/2B$, where $\lambda$ is the wavelength and $B$ is the baseline of $6.5$ m) of $93\times88$ mas$^{2}$ ($1.9\times1.8$ pc$^{2}$), $105\times101$ mas$^{2}$ ($2.1\times2.0$ pc$^{2}$), and $123\times116$ mas$^{2}$ ($2.5\times2.3$ pc$^{2}$) in the F380M, F430M, and F480M filters, respectively.

At all wavelengths ($3.8-4.8\,\mu$m), we find that the continuum dust emission has 
1) an extended component of $0.10" \times 0.25"$ ($\sim 2\times 5$ pc$^{2}$) at a PA $\sim -64^{\circ}$, 
b) a `North arc' feature extending $\sim4$ pc toward the north-east direction, and
c) a lack of IR emission (`Holes') in the north-east and south-west regions at $\sim0.4"$ ($\sim8$ pc) at a PA $\sim50^{\circ}$ from the nucleus. We identify the `North arc' as a real feature with $>16\sigma$ detection in the reconstructed images of all filters, as it persists after our bootstrapping analysis and independent image reconstructions (Methods section `Image reconstruction'). 
In addition, all images show an extended low-surface brightness emission at $>5$ pc, mostly along the directions of the narrow line region (NLR), and forming arcs around the `Holes'.

The AMI/JWST observations of Circinus improve the angular resolution of NIRCam/JWST by a factor of two---$0.08"$ vs. $0.14"$ at $4.3~\mu$m--, while removing the typical shape of the point spread function (PSF) of the JWST and filtering out the large-scale starlight emission from the host galaxy (the maximum recoverable scale is $\sim0.5"$ at $4.8~\mu$m). These benefits are clearly visualized in Figure \ref{fig:fig2}-top. Note that NIRCam/JWST observations will produce saturated images for Circinus and that the 8-m class single-dish observations using NACO/VLT and VISIR/VLT observations at $4.8~\mu$m \citep{Prieto2004} and $10.5~\mu$m \citep{Stalevski2017}, respectively are dominated by the large-scale starlight emission. Our AMI/JWST observations of Circinus show that the NIR emission within the central $10$~pc of Circinus is dominated by an extended emission feature physically linked to the AGN.


\section*{\textbf{The multi-phase components of the extended emission}} \label{sec:Multiphase_Images}

\begin{figure*}[ht!]
\includegraphics[width=\textwidth]{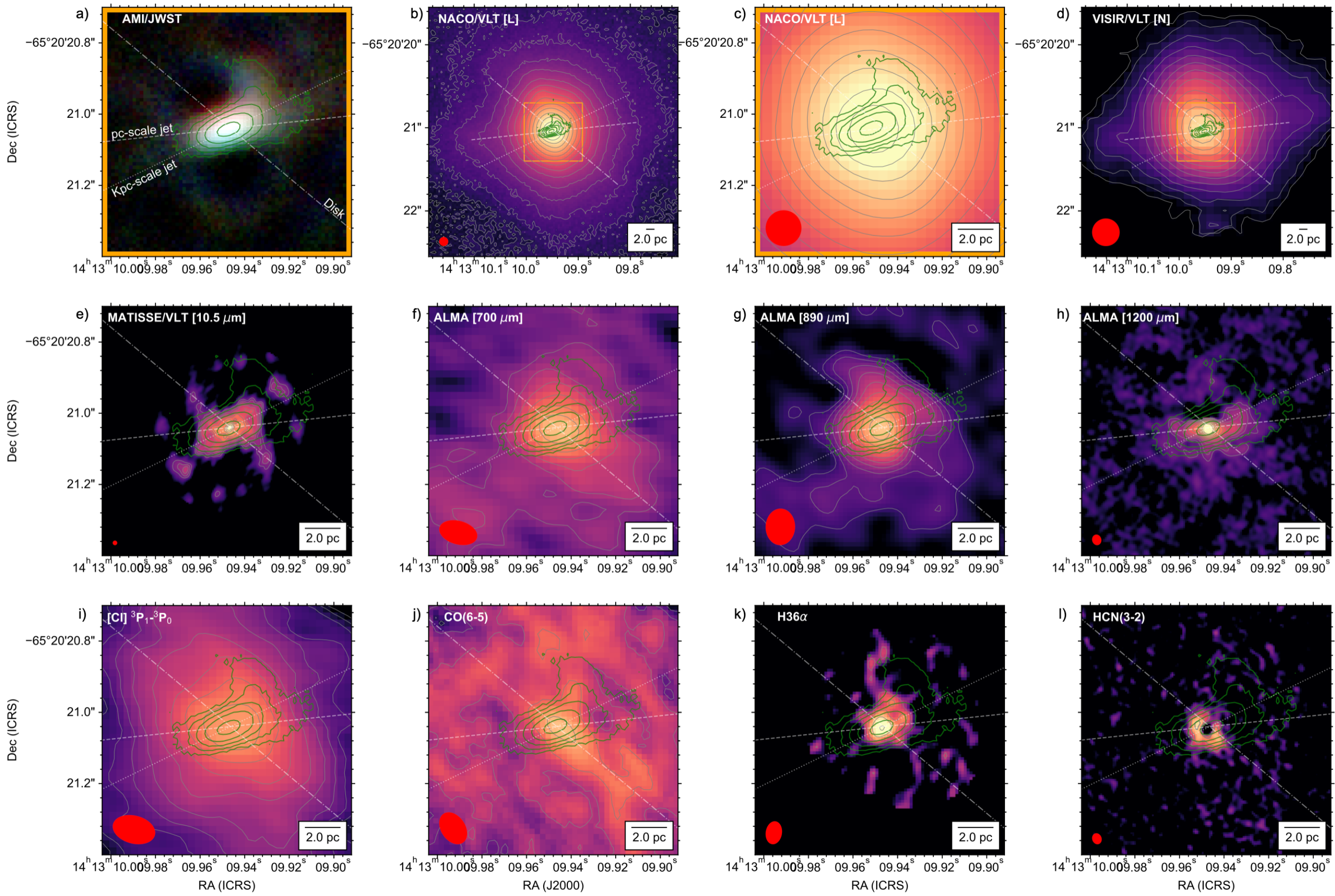}
\caption{\textbf{The multi-phases of the Circinus galaxy.}
\textbf{(a)} The RGB image of Circinus using the AMI/JWST observations and green contours of the $4.3~\mu$m AMI/JWST observations as shown in Fig. \ref{fig:fig1}. The orientations of the pc-scale jet (dashed line), kpc-scale jet (dashed line), and disk (dotted-dashed line) are shown.
\textbf{(b-d)} The single-dish continuum observations at $4.3~\mu$m (L-band) and $10~\mu$m (N-band) with the VLT \citep{Prieto2004} in the central $60\times60$ pc$^{2}$. The orange rectangle shows the $14\times14$ pc$^{2}$ within the $60\times60$ pc$^{2}$ large FOV images. 
\textbf{(e-h)} The interferometric continuum observations at $10.5~\mu$m with MATISSE/VLTI \citep{Isbell2022} and at $700$, $890$, and $1200~\mu$m with ALMA \citep{Izumi2023}.
\textbf{(i-l)} The interferometric observations \citep{Tristram2022,Izumi2023} of several gas tracers: [CI] $^{3}$P$_{1}$-$^{0}$P$_{1}$, CO(6-5), H36$\alpha$, and HCN(3-2). The gray contours show the morphology of the colorscale images in each panel. The beam of each observation (red ellipse) and a $2$ pc legend are shown.  \label{fig:fig2}}
\end{figure*}

We analyze the spatial correspondence of our measured $3.8-4.8\,\mu$m extended emission features with those arising from thermal and non-thermal continuum emission and molecular and ionized gas tracers (Methods section `Archival Observations'). We use observations with comparable, or better, angular resolution than the AMI/JWST observations. 
From the continuum emission observations (Fig.\,\ref{fig:fig2}-middle), the $3.8-4.8~\mu$m emission is cospatial, in extension and orientation, with the $8-13~\mu$m (N-band) emission observed with MATISSE/VLTI \citep{Isbell2022}. The $8-13~\mu$m extended emission has a dust temperature of $200-240$\,K along the direction of the kpc-scale jet, while it reaches a maximum dust temperature of $\sim270$\,K in the two East and West knots from the core and along the direction of the pc-scale jet.
In addition, the MATISSE/VLTI observations \citep{Isbell2023} at $3.7~\mu$m (L-band) show an unresolved source ($\sim0.3$ pc in diameter) and a resolved structure, $0.3\times0.6$ pc$^{2}$, at $\sim58^{\circ}$ highly offset from our extended emission (PA $\sim -64^{\circ}$) at $4.7~\mu$m (M-band). Due to the high angular resolution of all the MATISSE/VLTI observations (i.e., $4$ and $10$ mas in the LM and N bands, respectively), the `North arc' is filtered out.

The $700-1200~\mu$m continuum observations using ALMA show variations in morphology due to the change in the physical mechanism producing the emission \citep{Izumi2023}.
The $700~\mu$m observation is dominated by thermal dust continuum emission with an extension of $4\times10$ pc$^{2}$ at a PA $~\sim50^{\circ}$ \citep{Prieto2004,Prieto2010} (Methods Section `SED' and Fig. \ref{fig:app_fig6}). This structure is cospatial with the equatorial axis (i.e., disk) of the torus in Circinus and highly offset, $\Delta \rm{PA} = 66^{\circ}$, from our measured $3.8-4.8~\mu$m extended structure. The $700~\mu$m dust continuum emission along the disk fills the lack of emission observed in our $3.8-4.8~\mu$m AMI/JWST observations. Note also that the thickness, $\sim4$ pc, of the $700\,\mu$m dust emission spatially coincides with the long axis of the $3.8-4.8~\mu$m extended emission.
The $1200~\mu$m observation is dominated by non-thermal synchrotron emission arising from an unresolved nucleus ($\sim0.5$ pc), and a pc-scale jet extending $\sim4$ pc in diameter at a PA $= -84^{\circ}$ in the east-west direction. The eastern region of the synchrotron emission is cospatial with the dust continuum emission of our $3.8-4.8~\mu$m AMI/JWST observations. In the western region, the synchrotron emission shows an arc toward the northwest direction at $\sim3$ pc from the nucleus and is cospatial with the beginning of the `North arc' from our $3.8-4.8~\mu$m AMI/JWST observations. The two hottest knots within the central $0.5$\,pc measured in the N-band with MATISSE/VLT \citep{Isbell2022} are cospatial with the pc-scale jet at $1200~\,\mu$m.
The $890~\mu$m observations show the combined emission from dust and synchrotron with an unresolved core with extended emission in the northwest direction showing two arcs toward the north-east and south-west direction. The $890~\mu$m northern arc is slightly offset toward the inner side of the `North arc' at $3.8-4.8~\mu$m. In addition, the $890~\mu$m image also shows a lack of emission along the northeast region of the disk at PA $\sim50^{\circ}$. 

From the continuum observations, we conclude that our measured $2\times5$ pc$^{2}$ extended emission is highly offset, $\Delta \rm{PA} = 66^{\circ}$, from the equatorial axis of the obscuring disk around the nucleus with an east-west component parallel to the pc-scale radio jet at a PA $\sim-84^{\circ}$ extending $>5$ pc from the nucleus. This extended emission is also observed at $10\,\mu$m at 10s parsecs (Fig. \ref{fig:fig2}-d), which was attributed to the dusty cone edge directly radiated by the anisotropic radiation from the AGN \citep{Stalevski2017}. In addition, there is a lack of $3.8-4.8~\mu$m dust emission in our AMI/JWST observations, where the equatorial axis of the obscuring disk is present at $700\,\mu$m.

From the gas tracer observations (Fig. \ref{fig:fig2}-bottom), 
the [CI] $^{3}$P$_{1}$-$^{0}$P$_{1}$ and H36$\alpha$ show two arcs along the northwest direction, which have previously been identified as multi-phase gas outflows \citep{Izumi2023}. The molecular outflow in the [CI] $^{3}$P$_{1}$-$^{0}$P$_{1}$ observations is spatially coincidental with the `North arc' measured in our $3.8-4.8\,\mu$m AMI/JWST observations. The ionized outflow in the H36$\alpha$ is spatially coincidental with the base of the `North arc' in the $3.8-4.8\,\mu$m AMI/JWST observations. The CO(6-5) observations also show an emission structure cospatial with the first $\sim2$ pc of the `North arc' detected in the $3.8-4.8\,\mu$m wavelength range.

The HCN(3-2) observations show a $\sim2$ pc diameter disk at a PA $\sim50^{\circ}$, cospatial with the $700\,\mu$m dust continuum emission and the diameter along the short axis of the $3.8-4.8\,\mu$m extended emission. 
The CO(6-5) observations \citep{Tristram2022} are dominated by an extended emission $\sim6$ pc in diameter at a PA $\sim50^{\circ}$ highly offset, $\Delta \rm{PA} = 66^{\circ}$, from the measured dust continuum emission measured in the $3.8-4.8\,\mu$m AMI/JWST observations and cospatial with the dust continuum emission at $700\,\mu$m.
Weak molecular CO(6-5) emission is detected in the regions of lack of emission (`Holes') in the $3.8-4.8\,\mu$m wavelength range.

From the gas tracer observations, we conclude that 
the $3.8-4.8\,\mu$m dust continuum emission in the `North arc' is spatially coincidental with the [CI] $^{3}$P$_{1}$-$^{0}$P$_{1}$ and CO(6-5) molecular outflows and the H36$\alpha$ ionized outflow. The best spatial coincidence is with [CI] $^{3}$P$_{1}$-$^{0}$P$_{1}$ tracing diffuse atomic gas with a critical density of $n_{\rm{cr}}=3.7\times10^{2}$ cm$^{-3}$ \citep{Izumi2023}. The $3.8-4.8\,\mu$m `North arc' dust continuum emission may be a dusty phase entrained in the outflowing material at $\sim3$\,pc from the AGN.

We find no spatial correspondence of the $\sim 2 \times 5$ pc$^{2}$ extended component at a PA $\sim -64^{\circ}$ with any of the molecular or ionized gas tracers. This result may indicate that the $3.8-4.8\,\mu$m extended emission is mainly arising from directly radiated dust along the funnel of the obscuring disk and NLR. As mentioned above, the $>5$ pc dust emission along the east-west direction can be attributed to directly radiated dust by the pc-scale jet \citep{Stalevski2017}.

We find that the region with a lack of $3.8-4.8\,\mu$m dust emission (`Holes') is spatially coincidental with CO(6-5) molecular emission, the dust continuum emission at $700\,\mu$m, and the $\sim2$ pc disk observed at HCN(3-2). The HCN(3-2) is cospatial with the maser disk \citep{Greenhill2003}. The lack of $3.8-4.8\,\mu$m  dust emission in the northern and southern regions at $\sim 8$ pc at a PA $\sim50^{\circ}$ from the nucleus may be caused by obscuration effects due to the optically thick obscuring disk around the AGN.


\section*{\textbf{Origin of the central continuum emission}} \label{sec:SED}

\begin{figure*}[ht!]
\includegraphics[width=\textwidth]{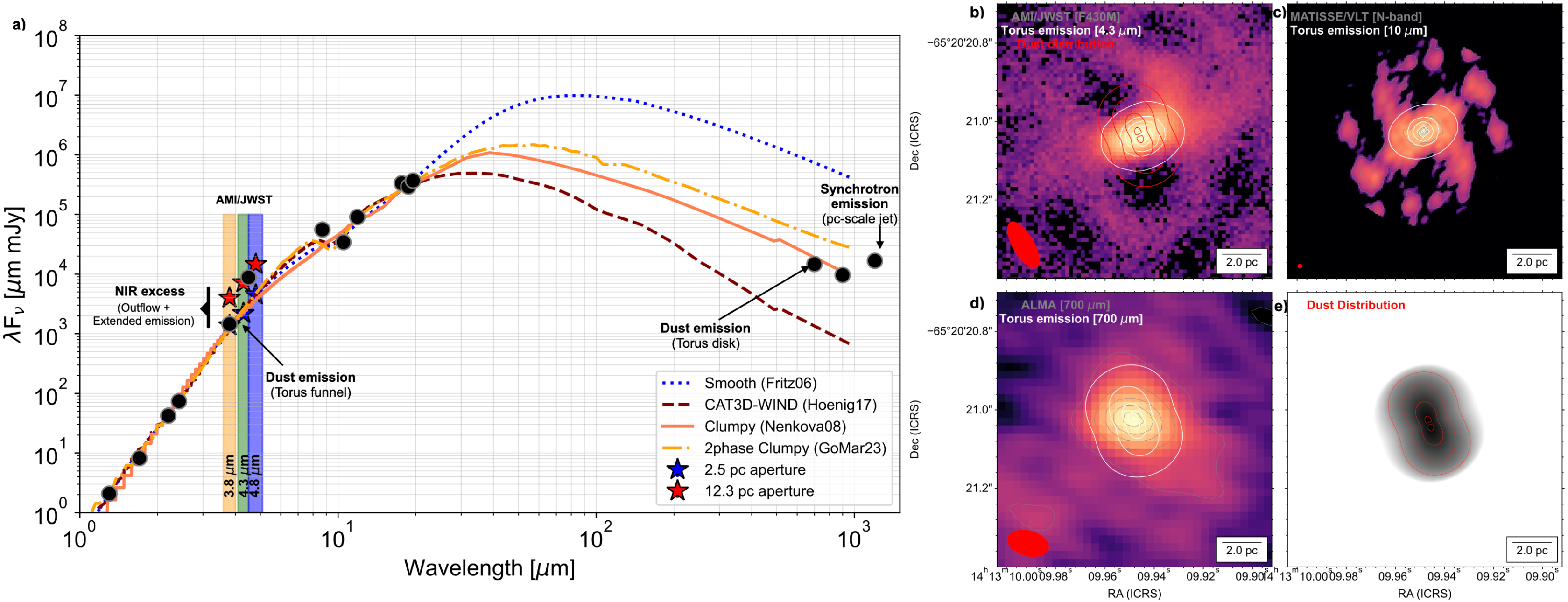}
\caption{\textbf{The \textsc{clumpy} torus model describes the $1-1000\,\mu$m SED and morphological changes of the dust emission of Circinus.}
\textbf{a)} The $1-1000\,\mu$m photometric measurements from the literature (black dots) and our AMI/JWST observations at a $2.5$ pc aperture (blue star) and at a large aperture of $12.3$ pc (red star) of the central emission. The bandwidths of the AMI/JWST filters are shown as shadowed regions. The best-fits of the several torus models are shown. Labels indicate the dominant emitting and physical components.  
\textbf{(b-d)} Dust emission at $4.3\,\mu$m, $10\,\mu$m, and $700\,\mu$m as shown in Fig. \ref{fig:fig2} with the synthetic dust emission (white contours) using \textsc{Hypercat} with the the best-fit \textsc{clumpy} torus model. The dust mass distribution of the best-fit \textsc{clumpy} torus model is shown in red contours and panel \textbf{(e)}.
 \label{fig:fig3}}
\end{figure*}

To quantify the dust temperature and relative contribution of the $3.8-4.8\,\mu$m observed morphological structures, we perform a photometric analysis of the central $10\times10$ pc$^{2}$ (Methods section `Photometry'). Using a 2D Gaussian profile to fit the central extended emission, we estimate that $87^{+5}_{-7}$\% of the total flux arises from the central $2\times5$ pc$^{2}$ at a PA $=-70^{\circ}$. We fit a blackbody function to the photometric measurements and estimate a characteristic dust temperature of $450\pm10$\,K. Using $\sim0.1"-0.4"$ aperture photometry from single-dish telescopes, a characteristic dust temperature of $300$ K was measured \citep{Prieto2004}. This component was attributed to a $\sim2$ pc diameter dusty torus obscuring the AGN. The lower dust temperature may be due to the starlight contaminated by the host galaxy.

After extracting the central extended emission, we estimate that only $\sim1$\% of the total emission arises from the `North arc'. This dust component has a characteristic dust temperature of $730\pm20$\,K. Most of the extended emission, $12^{+4}_{-6}$\% of the total flux, arises from dust located at $>5$ pc from the core, mostly in the east-west direction at a PA $\sim-84^{\circ}$ (cospatial with the pc-scale jet) with a characteristic dust temperature of $420\pm20$\,K. These results indicate that the `North arc' is an independent dust component not associated with the central extended emission. The $>5$ pc extended emission may be dust in the NLR directly radiated by the pc-scale jet and/or AGN--a continuation of the central elongated emission extending to 10s of pc scales \citep{Stalevski2017}.

To identify the dominant physical component (i.e., disk vs. wind) of the central elongated emission, we analyze the $1-1000\,\mu$m SED by fitting several AGN torus models (Methods section `Torus models'). The SED was constructed using photometric measurements with a spatial resolution comparable to those by AMI/JWST (Figure \ref{fig:fig3}; Methods section `SED'). We use torus-only models with smooth \citep[`Smooth';][]{Fritz2006}, clumpy \citep[`\textsc{clumpy}';][]{Nenkova2008a,Nenkova2008b}, and the 2-phase clumpy torus with inclusion of dust grain sizes \citep[`2phase clumpy'][]{GM2023}, and the torus+wind model \citep[`CAT3D-WIND'][]{Hoenig2017}. For all models, we fixed the inclination of the disk to be edge-on, $i=90^{\circ}$ as shown by the maser disk \citep{Greenhill2003}, and we include a dust screen as a free parameter.

Including our new AMI/JWST photometric measurements, we find that the torus-only models are statistically preferred to describe the $1-1000\,\mu$m SED of the central $2\times5$ pc$^{2}$ of Circinus (Fig. \ref{fig:fig3}, Method section `Table 2'). The largest differences between models arise in the sub-mm wavelength regime (i.e., cold dust component) and the $9.7\,\mu$m silicate feature. Note that the sub-mm regime traces the bulk of the dust mass in the torus \citep{ELR2018,HypercatI,HypercatII}, which provides a crucial photometric measurement to characterize the torus models. The $1-20\,\mu$m SED (including the silicate feature) of the central $2\times5$ pc$^{2}$ statistically prefers the `2phase clumpy' model, followed by the smooth torus model. The $1-1000\,\mu$m SED of the central $2\times5$ pc$^{2}$ statistically prefer the \textsc{clumpy} torus models, followed by the CAT3D-WIND model. The \textsc{clumpy} torus model underpredicts the 9.7 silicate feature at small scales, except for the large $8$ pc aperture, indicating that a dust screen may be present along the LOS. This is also found in the MATISSE/VLTI observations, showing that a uniform screen of foreground absorption of $A_{V}=28.5$ mag. is required to reproduce the silicate feature. The CAT3D-WIND models underpredict the sub-mm dust emission. This is because most of the dust is relocated above and below the disk (i.e., polar direction), leaving a small and compact dusty disk along the equatorial axis. Most of the SED is then dominated by the hot and warm dust in the base of the outflow. The Smooth models overpredict the sub-mm dust emission. This is because the torus is an optically and geometrically thick and monolithic dusty torus. The 2-phase clumpy models overpredict the sub-mm emission. This is because the interclump dust component provides an extra cold emission component to that from the clumps.

To analyze the morphological changes of the continuum thermal emission of the torus, we compute synthetic surface brightness and cloud distribution images using the radiative transfer code Hypercubes of AGN Tori \citep[\textsc{HyperCAT};][]{HypercatI,HypercatII}. \textsc{HyperCAT} uses the \textsc{clumpy} torus models with any combination of torus model parameters to generate physically scaled and flux-calibrated 2D images of the dust emission and distribution for a given AGN. We use the statistically preferred \textsc{clumpy} torus models model of the $1-1000\,\mu$m central $2\times5$ pc$^{2}$ with an optical depth per cloud of $\tau_{V}\sim20$ with an average of $\sim4$ clouds along the radial equatorial plane. We use a distance of $4$ Mpc, and a tilt angle on the plane of the sky of $50^{\circ}$  cospatial with the orientation along the lack of emission in our AMI/JWST observations. The \textsc{HyperCAT} images are then smoothed and pixelated to match the $3.8-4.8\,\mu$m AMI/JWST, $10.5\,\mu$m MATISSE/VLTI, and $700\,\mu$m ALMA observations. In Fig. \ref{fig:fig3}, we show that the observed morphological changes in the distribution of dust emission from the IR to the sub-mm wavelengths can be reproduced by the \textsc{clumpy} torus model with a size of $5 \times 3$ pc$^{2}$. 

The central $2\times5$ pc$^{2}$ extended component in the $3.8-10\,\mu$m wavelength range can be attributed to thermal emission from directly radiated dust along the funnel of the torus. The lack of $3.8-10\,\mu$m emission perpendicular to the extended component of central $2\times5$ pc$^{2}$ in the $3.8-10\,\mu$m wavelength range can be attributed to optically thick dust along the equatorial axis of the disk. This component is then observed at $700\,\mu$m, which traces the cold dust along the equatorial axis of the disk--co-spatial with the bulk of the dust distribution. 

We study the dependency of model fitting with the photometric aperture size (See Appendix Figure \ref{fig:app_fig6}). We find that the $1-1000\mu$m SED statistically prefers the \textsc{clumpy} torus models for apertures encompassing the central $\sim3\,$pc extended emission. Using a large aperture of $8\,$pc that contains the North Arc and the emission along the NLR, the smooth torus models best describe the $1-1000\,\mu$m SED with the clumpy torus models underestimating the NIR and sub-mm fluxes, and the 2-phase clumpy model overestimating the sub-mm fluxes. The $9.7\mu$m feature is well described by all models using large-aperture SED. These results imply that there is a diffuse extended emission at $>4$\,pc-scales, not associated with the central torus, that needs to be accounted for. This NIR excess arising from the outflow and extended emission in the NLR (Fig. \ref{fig:app_fig3}). This component may be accounted for by the CAT3D-WIND, which is the next preferred model for the $1-1000\mu$m SED (Methods section Table \ref{tab:app_table2}).

In conclusion, the $3-5\,\mu$m emission excess mainly ($87$\%) arises from dust within the $5$ pc around the AGN associated with the funnel of the postulated torus. This central extended emission has no spatial correspondence with any of the molecular and ionized outflowing gas at a similar, or better, spatial resolution. The morphological changes of the dust emission are consistent with those expected by a \textsc{clumpy} torus with the dust distribution located on a disk of $\sim5\times3$ pc$^{2}$ at a PA$\sim50^{\circ}$. Thus, most of the mass reservoir is located in the form of an accreting dusty disk that feeds the SMBH of Circinus. 

The anisotropic emission measured along the east-west direction dominated by direct radiation from the pc-scale jet requires an additional component not associated with the torus \citep{Stalevski2017,Stalevski2019}. This component accounts for $\sim12$\% of the $3.8-4.8\,\mu$m emission at $>5$ pc scales, which we attribute to dust in the NLR directly irradiated dust by the AGN and by the pc-scale jet. Recent studies (Campbell et al. 2025, Lopez-Rodriguez et al. 2025a, submitted) using MIRI/MRS/JWST observations of $6$ nearby AGN have shown that the $39-450$\,pc MIR extended emission accounts for $<40$\% of the total $10\,\mu$m emission. This MIR emission is attributed to dust located in molecular clouds and/or shocks in the NLR heated by outflows and/or the central AGN.

Our observations show that $<1$\% of the $3.8-4.8\,\mu$m emission arises from dust entrained in the multi-phase outflow labeled as `North Arc'. This structure has been spatially correlated with molecular and ionized gas at similar, or better, spatial resolution. The `North arc' may be dust entrained in the outflow, which requires an additional physical component to that of the dusty torus or dust in the NLR. Indeed, an outflowing material has been modeled to explain the anisotropic radiation \citep{Stalevski2017,Stalevski2019}.

We show that the combination of the $3-10\,\mu$m and $700-1200\,\mu$m imaging observations is critical to disentangle the dust emission components of the central $10$ pc around AGN. In addition, complementary molecular and ionized gas is required to spatially correlate dust entrained in outflows. Our JWST interferometric observations open a new window to disentangle the accreting and outflowing dusty components in the central $10$ pc of nearby AGN without the effect of the JWST's PSF, saturation in the imaging mode, and large-scale starlight component of the host galaxy.

\begin{acknowledgments}
This work is based on observations made with the NASA/ESA/CSA James Webb Space Telescope. The data were obtained from the Mikulski Archive for Space Telescopes at the Space Telescope Science Institute, which is operated by the Association of Universities for Research in Astronomy, Inc., under NASA contract NAS 5-03127 for JWST. These observations are associated with program \#4611.
Support for program \#4611 was provided by NASA through a grant from the Space Telescope Science Institute, which is operated by the Association of Universities for Research in Astronomy, Inc., under NASA contract NAS 5-03127.
E.L.-R. thanks support by the NASA Astrophysics Decadal Survey Precursor Science (ADSPS) Program (NNH22ZDA001N-ADSPS) with ID 22-ADSPS22-0009 and agreement number 80NSSC23K1585. J.S.-B. acknowledges the support received by the UNAM DGAPA-PAPIIT project AG 101025.
\end{acknowledgments}

\begin{contribution}

E.L.R. led the JWST proposal, data reduction, image reconstruction, scientific analysis, and manuscript writing and editing the full manuscript. 
J.S.-B performed the image reconstruction, interferometric analysis, and writing the image reconstruction section.
O.G.-M performed the torus models fitting and scientific interpretation.
R.N. supported the JWST proposal, the HyperCAT images, and scientific interpretation.
R.M.L. supported the JWST proposal, JWST observations, and image reconstruction.
D.T. supported the JWST proposal, JWST observations, and AMI data reduction.
I.G.M. supported the emission line contamination analysis and scientific interpretation.
J.H.G. supported the JWST proposal and JWST observations.
M.J.H. supported the JWST proposal.


\end{contribution}

%



\appendix

\section{Methods} \label{App:Methods}

\begin{figure*}[ht!]
\includegraphics[width=\textwidth]{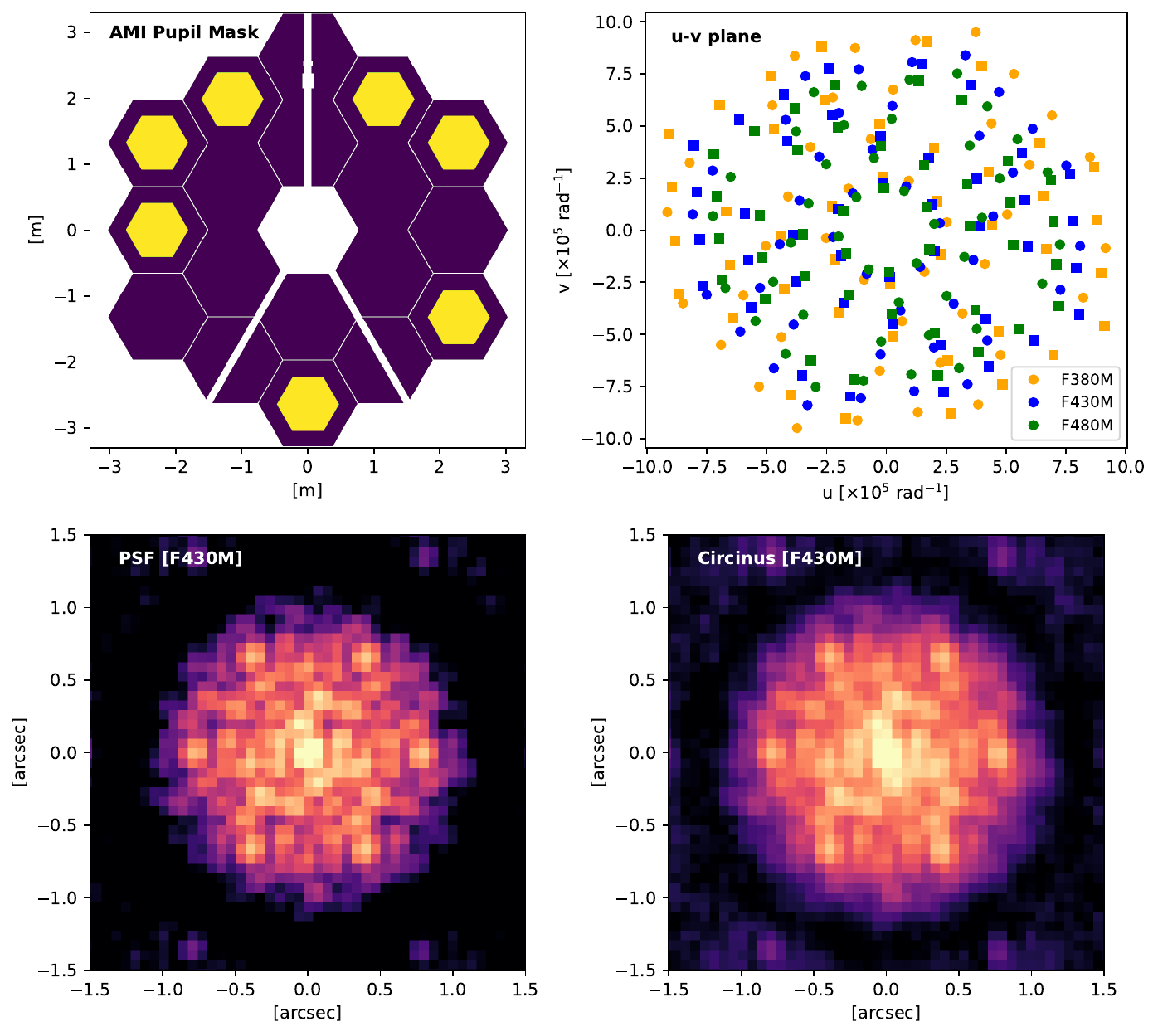}
\caption{\textbf{Example of AMI/JWST observations.} 
The 7-hole non-redundant NIRISS AMI pupil mask (yellow hexagons) over the JWST primary mirror (top left).
The u-v coverage (top right) of the Circinus observations in the F380M (orange), F430M (blue), and F480M (green) filters. The u-v coverage for the first (circles) and second (squares) epochs is shown.
The central $3\arcsec\times3\arcsec$ ($45\times45$ px$^{2}$) interferogram image on the $80\times80$ px$^{2}$ SUB80 array with the F430M filter of the standard star HD119164 (bottom left) and Circinus (bottom right). The FOV of the Circinus observations is $90\times90$ pc$^{2}$. Both interferogram images are normalized to peak. Compared with the standard star, the Circinus observations show an elongated feature within the central $0.5"$ and a large-scale extended emission.
 \label{fig:app_fig1}}
\end{figure*}

\begin{figure*}[ht!]
\includegraphics[width=\textwidth]{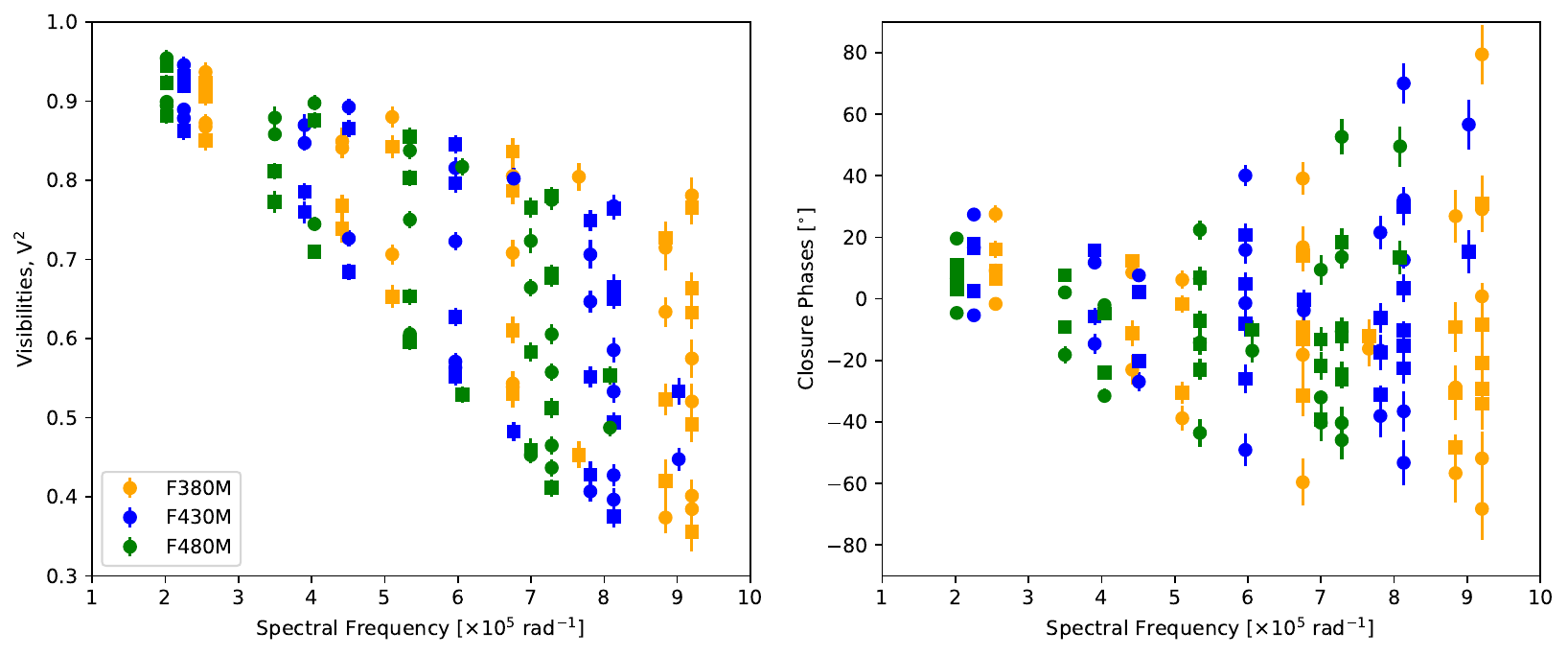}
\caption{\textbf{Interferometric observables of Circinus.} 
The calibrated visibilities (left) and closure phases (right) for the Circinus observations in the F380M (orange), F430M (blue), and  F480M (green) filters for the first (circles) and second (squares) epochs.
 \label{fig:app_fig2}}
\end{figure*}

\textbf{Observations.} Circinus was observed (ID: 4611; PI: Lopez-Rodriguez, E.) on 20240715 and 20250427 using the Aperture Masking Interferometry \citep[AMI;][]{AMI2023} mode of the Near Infrared Imager and Slitless Spectrograph \citep[NIRISS;][]{NIRISS2012} instrument on the JWST. We performed observations of Circinus and a standard star, HD119164, using the F380M ($\lambda_{\rm{c}} = 3.827\,\mu \rm{m}, \Delta\lambda = 0.21\,\mu \rm{m}$), F430M ($\lambda_{\rm{c}} = 4.326\,\mu \rm{m}, \Delta\lambda = 0.20\,\mu \rm{m}$), and F480M ($\lambda_{\rm{c}} = 4.817\,\mu \rm{m}, \Delta\lambda = 0.30\,\mu \rm{m}$) filters. For all observations, the AMI pupil mask (Fig. \ref{fig:app_fig1}), the SUB80 ($80\times80$ px$^{2}$) array, and the NISRAPID readout pattern were used with a pixel scale of $65$ mas px$^{-1}$ and a readout of 75.44 ms. To avoid signal limit ($26,000$ e$^{-}$) of the standard star during the observations, we used a setup with 145, 338, and 190 integrations and 4, 4, and 5 groups in the filter sequence of F480M, F380M, and F430M, respectively. This sequence is used to optimize the direction of rotation of the filter wheel and prolong the life of the mechanism. For both Circinus and standard star, the total/effective exposure times are 102/76 s,  57/72 s, and 44/33 s in the F380M, F430M, and F480M filters, respectively. The Stare mode without a dither pattern was used. These are the first two observations of a set of three from this JWST program to rotate the uv-plane of the observations. The V3 position angles are $4^{\circ}$ and $87^{\circ}$ for the first and second epoch of observations. The uv-plane has rotated by $87^{\circ}-4^{\circ}\sim83^{\circ}$ (Fig. \ref{fig:app_fig2}), which agrees with the $\sim90^{\circ}$ with a $10^{\circ}$ margin offset between both epochs requested for these observations.

\textbf{Data Reduction.} We processed the NIRISS AMI observations using the JWST Calibration pipeline (version 1.15.1; CRDS version 11.17.26) and the CRDS context \textsc{jwst\_1258.pmap}. We followed the standard NIRISS AMI data reduction recipe for stages 1 to 4. Stage 1 (\textsc{calwebb\_detector1}) produces corrected count rate images (`rate' and `rateints' files) after performing several detector-level corrections. Stage 2 (\textsc{calwebb\_image2}) produces calibrated exposures (`cal' and `calints' files), where we skip the photometric calibration (photom=False) and resampling (resample=False) of images. These steps produce the interferogram image in units of digital counts shown in Fig. \ref{fig:app_fig1}. Stage 3 (\textsc{ami\_analyze}) is a specific pipeline step for the AMI observations. This step produces the interferometric observables (`ami-oi' files) after computing fringe parameters for each exposure producing an average fringe result of the full observations. The uv coverages of the observations for both epochs are shown in Fig. \ref{fig:app_fig1}. Stage 4 (\textsc{ami\_normalize}) produces (`amimorn-oi' files) the final normalized interferometric observables after correcting the science target using the reference standard star. The normalized and calibrated visibilities, V$^{2}$, and closure phases for both epochs are shown in Fig. \ref{fig:app_fig2}.

Additionally, we obtained calibrated interferometric observables using SAMPip \citep{Sanchez-Bermudez_2022}. This software uses a fringe-fitting routine to look for the amplitude and phase solutions that recover the structure of the interferogram, considering the non-redundant mask geometry of NIRISS/JWST. Each frame in the data cubes is fitted individually, and the final squared visibilities and closure phase values are averaged per data cube with their corresponding standard deviations. The observables from the science data cubes are corrected by the instrumental transfer function using the standard star HD\,119164. The calibrated observables are stored in standard OIFITS files \citep{Pauls_2005} for posterior analysis.

\textbf{Image reconstruction.} We reconstructed the Circinus images at each filter using SQUEEZE\footnote{GitHub repository of SQUEEZE: \url{https://github.com/fabienbaron/squeeze}} \citep[version 2.7; ][]{squeeze2010}.
This algorithm has successfully been used to reconstruct the NIRISS AMI observations of the Wolf-Rayet, WR 137 \citep{Lau2024}. SQUEEZE reconstruction image algorithm uses a Markov Chain Monte-Carlo (MCMC) approach to explore the imaging probability space using the interferometric observables with its associated uncertainties. Using the SAMPip outputs, SQUEEZE images were recovered using a pixel grid of $129 \times 129 $ px$^{2}$ (FOV $= 1.29 \times 1.29$ arcsec$^{2}$), with a pixel scale of 10 mas. For the reconstruction, we used two regularization functions, a Laplacian and the L0-norm, with the following hyperparameter values $\mu_{\mathrm{La}} = 500$ and $\mu_{\mathrm{L0}} = 0.2$, respectively. With these parameters, the reconstructions converged with $\chi_{\nu}^2$ close to unity. To characterize the signal-to-noise ratio (SNR) of the images, we recovered $100$ images per data cube with different samples of the observed uv-plane. For this procedure, we randomly sampled the uv frequencies of the interferometer by changing their weights, at the time we keep the total number of uv points constant. Finally, we averaged the $100$ different reconstructions per wavelength. We computed the dirty beam of each filter shown in Fig. \ref{fig:app_fig4}. The angular resolutions in the reconstructed images of Circinus are $93\times88$ mas$^{2}$ ($1.9\times1.8$ pc$^{2}$), $105\times101$ mas$^{2}$ ($2.1\times2.0$ pc$^{2}$), and $123\times116$ mas$^{2}$ ($2.5\times2.3$ pc$^{2}$) in the F380M, F430M, and F480M filters, respectively, which corresponds to the theoretical angular resolution of $\lambda/2B$ by the interferometric observations. 

To estimate the validity of the reconstructed images compared with the calibrated observables, we computed the synthetic interferometric observables from each one of the recovered images per wavelength. The mean value and the standard deviation of the synthetic observables are shown in Fig. \ref{fig:app_fig4}. It can be observed that all data points from the images are consistent with the data within $1\sigma$. Similarly, to estimate the statistically significant features with SNR above the noise level in the images, we estimated their noise statistics ($\mu_{\mathrm{noise}}$, $\sigma_{\mathrm{noise}}$) using all the pixels values outside a box of $40 \times 40$ px$^{2}$ ($8\times8$ pc$^{2}$) centered in the image. The interferometric observables of those filtered images are consistent within $1\sigma$ of the reported synthetic observables, allowing us to trust the significance of the recovered morphology.

\begin{figure*}[ht!]
\includegraphics[width=\textwidth]{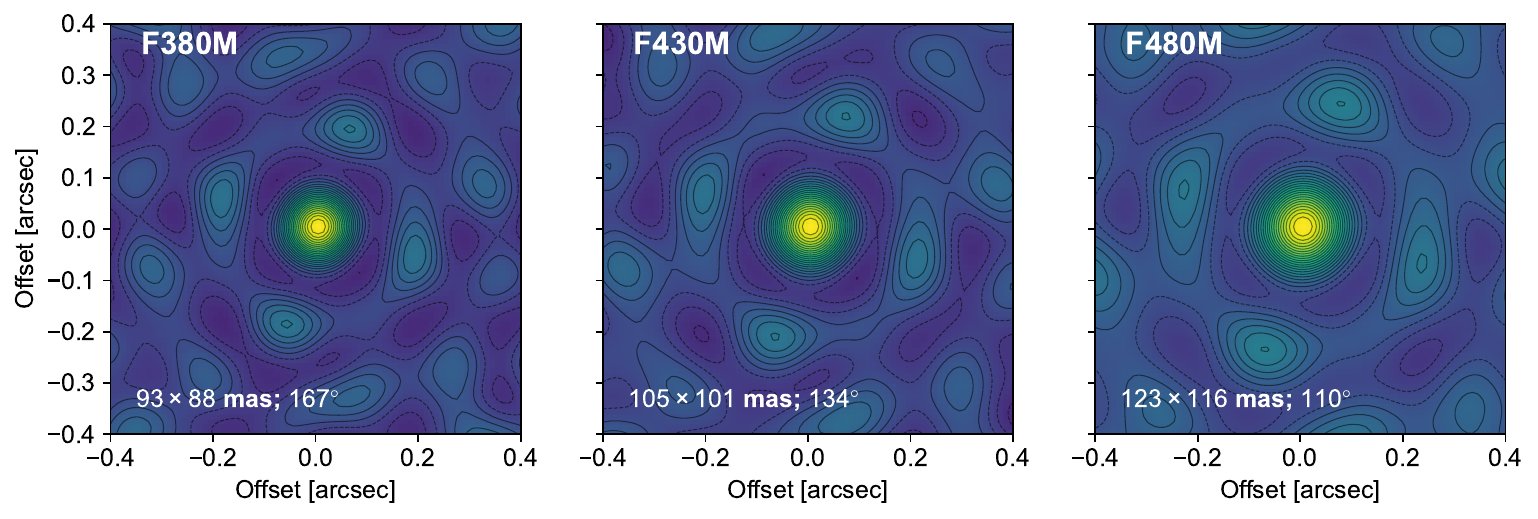}
\caption{\textbf{Dirty Beams.} The dirty beams of the F380M (left), F430M (middle), and F480M (right) filters. Dirty beams of the combined epochs. The FWHMs and PA of the beams are shown at the bottom left of each panel. Contours start at $-0.2 \times I_{\rm{peak}}$ and increase in steps of $0.05$, where $I_{\rm{peak}}$ is the peak of the beam scaled to unity.
 \label{fig:app_fig3}}
\end{figure*}

\begin{figure*}[ht!]
\includegraphics[width=\textwidth]{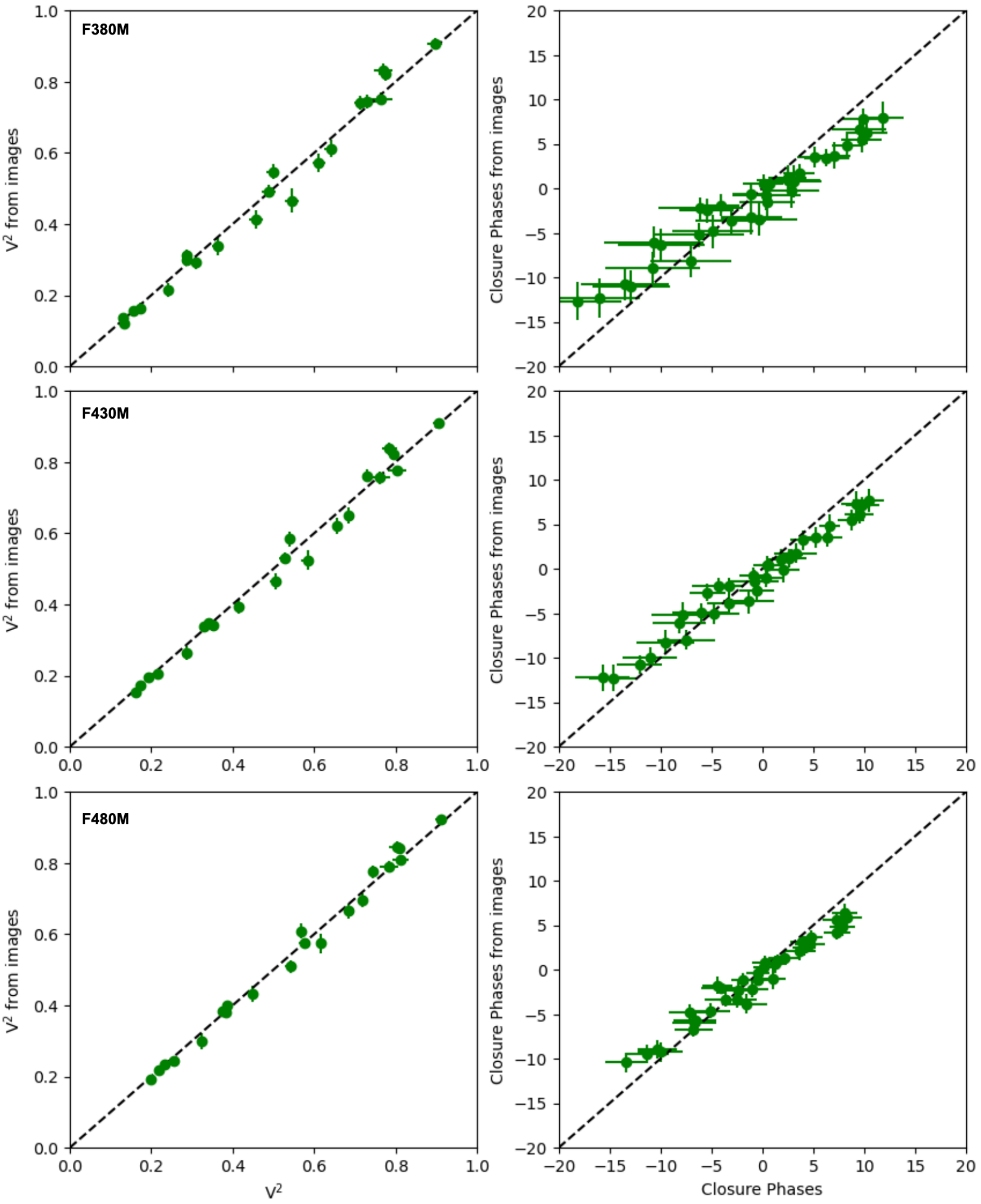}
\caption{\textbf{Comparison between observables and synthetic interferometric observables.}
The amplitudes (left) and closure phases (right) for the F380M (top), F430M (middle), and F480M (bottom) filters using the bootstrapping analysis described in `Image reconstruction'.
 \label{fig:app_fig4}}
\end{figure*}

\textbf{WCS correction.} The reconstructed images do not have a world coordinate system (WCS) associated with them. However, the interferogram pattern (Fig. \ref{fig:app_fig1}) has the WCS from the JWST observations. Thus, we use the sky coordinates of the peak from the interferogram pattern as the sky coordinates of the peak pixel from the reconstructed images at each filter. Here, we assume that the peak pixel of the reconstructed image is the position of the AGN, which dominates the IR emission of the object in both the interferogram pattern and the reconstructed images. A small WCS shift was then performed to align the AMI/JWST observations with the peak emission of the $1200\,\mu$m ALMA observations (Fig. \ref{fig:fig2}).

\textbf{Flux calibration.} The standard star HD119164 ($F_{12\,\mu m} = 1.2$\,Jy) was observed immediately after the science object using the same configuration as those for the Circinus galaxy. We took observations of the same standard star as previously used by the interferometric observations of Circinus with MATISSE/VLTI at L, M, and N-bands \citep{Isbell2022,Isbell2023}. The standard star serves to perform the flux calibration and final visibilities of the science object. The flux calibration of the final reconstructed image of the science object was computed such as:

\begin{equation}\label{eq:fluxcal}
F_{\rm{obj}}^{\rm{cal}}(\lambda)\,[\rm{Jy}] = F_{\rm{obj}}^{\rm{norm}}(\lambda) \times \frac{F_{\star}(\lambda)\,[\rm{Jy}]}{F_{\star}^{\rm{T}}(u=0,v=0,\lambda)\,[\rm{ADU}]} 
\times F_{\rm{obj}}^{\rm{T}}(u=0,v=0,\lambda)\,[\rm{ADU}]\,,
\end{equation}
\noindent
where $ F_{\rm{obj}}^{\rm{norm}}(\lambda)$ is the normalized reconstructed image of the science object with a total flux equal to unity, $F_{\star}(\lambda)$ is the total flux of the standard standard star in units of Jy, $F_{\star}^{\rm{T}}(u=0,v=0,\lambda)$ is the total flux of the zero-baseline of the standard star in units of counts (i.e., ADU: analog diginal unit), and $F_{\rm{obj}}^{\rm{T}}(u=0,v=0,\lambda)$ is the total flux of the zero-baseline of the science object in units of counts (i.e., ADU). All these fluxes are at a given wavelength, $\lambda$.

$F_{\star}(\lambda)$ was estimated using the spectral type, G8II, of the standard star, scaled to have a flux of $1.2$\,Jy at $12\,\mu$m \citep{Isbell2022,Isbell2023}. Then, we estimated the total flux of the standard star within the bandpass\footnote{The NIRISS throughputs can be found at \url{https://jwst-docs.stsci.edu/jwst-near-infrared-imager-and-slitless-spectrograph/niriss-instrumentation/niriss-filters\#NIRISSFilters-NIRISSsystemthroughput}} of the NIRISS/AMI filters to be $8.55$, $7.09$, and $5.95$ Jy at F380M, F430M, and F480M, respectively. 
$F_{\star}^{\rm{T}}(u=0,v=0,\lambda)$ was estimated using the total flux of zero-baseline from the image of the mirrored Hermitian counterparts in this uv-plane coverage, or Modulation Transfer Function \citep[MTF; see fig. 1 by][]{AMI2023}. The zero-baseline contains the total flux of the observations. We computed the total flux from the central peak of the MTF image using two methods. First, we perform aperture photometry with a radius of $3.5$ pixels. Second, we fit a 2D Gaussian profile with two free parameters: the amplitude and the FWHM, which is assumed to be axisymmetric. We estimate that the aperture photometry (i.e., first method) misses $\sim12-22$\% of the flux arising from the winds of the 2D Gaussian profile. We use the total flux of the zero-baseline estimated with the 2D Gaussian fitting profile. NIRISS AMI mode has a photometric calibration uncertainty\footnote{NIRISS AMI photometric calibration: \url{https://jwst-docs.stsci.edu/depreciated-jdox-articles/jwst-data-calibration-considerations/jwst-calibration-uncertainties\#JWSTCalibrationUncertainties-Photometriccalibration.10}} of $5$\% in the F380M and F430M filters and $8$\% in the F480M filter.

\textbf{Emission line contribution.} To estimate the potential contribution of spectral features within the filters, we use synthetic photometry on both the observed spectra and the feature-free continuum spectra of local AGN. First, we establish a baseline representing the continuum emission from the central spectrum by fitting feature-free continuum anchor points with straight lines \citep[e.g.,][]{Bernete24b}. Using the fitted baseline, we then perform synthetic photometry for the NIRISS imaging bands by convolving the spectra with the corresponding filter transmission curves\footnote{\url{http://svo2.cab.inta-csic.es/svo/theory/fps/index.php?mode=browse\&gname=JWST\&gname2=NIRISS&asttype=}}, as in \cite{Bernete24c} (see also \citealt{Donnelly25}). The main features contributing to the F380M, F430M, and F480M filters include gas-phase and icy molecular bands such as the $^{12}$CO ($\sim4.45-4.95\,\mu$m)  molecular gas-phase absorption band and the $4.27\,\mu$m stretching mode of the CO$_2$ ice \cite[e.g.,][]{Bernete24b,Gonzalez-Alfonso24,Pereira-Santaella24}.  We utilize local AGN (NGC\,3256 and NGC\,7469) from the Director's Discretionary Early Release Science Program \#1328 (PIs: L. Armus \& A. Evans) to calculate the fractional contribution of the continuum to the photometry of type 1 and type 2 AGN. For type 2 AGN, we find a continuum contribution of $94$\%, $71$\%, and $84$\% in the F380M, F430M, and F480M filters, respectively. Note that these continuum contributions should be considered a lower limit, as the lines are known to be stronger in luminous IR galaxies, such as NGC\,3256, which was used in this estimation. In the case of the type 1 NGC\,7469, the continuum dominated the emission in all the filters used in this work.

\textbf{Archival observations.} For our imaging analysis, we use the following archival observations. 
NACO/VLT images at L'-band ($3.8\,\mu$m, $\Delta\lambda=0.62\,\mu$m) with an FWHM of $0.12"$ \citep{Prieto2004}. 
VISIR/VLT at N-band ($10.5\,\mu$m; $\Delta\lambda=0.01\,\mu$m) with an FWHM of $0.3"$ \citep{Stalevski2017}.
MATISSE/VLTI at $10.5\,\mu$m with an FWHM of $10$ mas \citep{Isbell2022}. 
Dust continuum emission at $700\,\mu$m and non-thermal emission at $1200\,\mu$m using ALMA with beam sizes of $107\times64$ mas$^{2}$ at a PA $=74^{\circ}$ and $27\times24$ mas$^{2}$ and PA $=15^{\circ}$, respectively \citep{Izumi2023}.
Dust and radio continuum emission at $890\,\mu$m with a beam size of $100\times80$ mas$^{2}$ and PA $=-1.7^{\circ}$ (Lopez-Rodriguez et al., in prep.).
[CI] $^{3}$P$_{1}$-$^{0}$P$_{1}$ with a beam size of $119\times76$ mas$^{2}$ and PA $=75^{\circ}$, H36$\alpha$ with a beam size of $62\times43$ mas$^{2}$ and PA $=-8^{\circ}$, and  HCN(3-2) with a beam size of $29\times24$ mas$^{\circ}$ and PA $=20^{\circ}$ \citep{Izumi2023}.
CO(6-5) with a beam size of $95\times66$ mas$^{2}$ and PA $=34^{\circ}$ \citep{Tristram2022}.

\begin{figure*}[ht!]
\includegraphics[width=\textwidth]{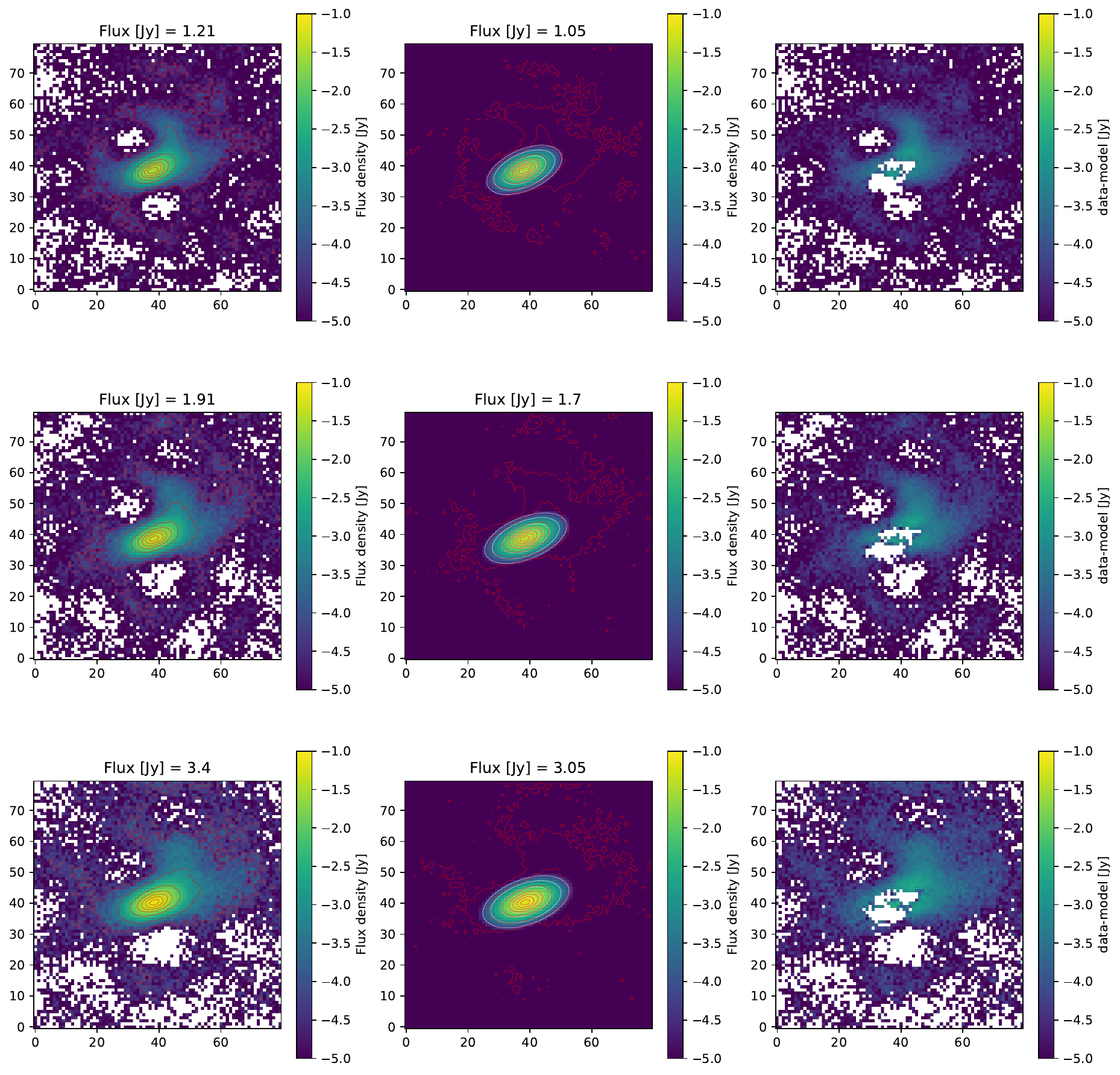}
\caption{\textbf{The results of the 2D Gaussian fitting to the extended emission in each of the AMI/JWST observations.} The AMI/JWST observations (first column), best-fit 2D Gaussian model (middle column), and residuals (third column) for the F380M (first row), F430M (middle row), and F480M (bottom row) filters. The red contours show the flux density of the observations, the white contours show the fluxes of the 2D Gaussian profiles. All contours have the same fractional levels from the peak pixel at [$10^{-3}$, $10^{-2}$, $10^{-1}$, $0.3$, $0.5$, $0.7$, $0.9$]. The axes are in pixels with a pixel scale of $10$ mas. The total flux densities within the FOV of $14\times14$ pc$^{2}$ are shown.
 \label{fig:app_fig5}}
\end{figure*}

\textbf{Photometry.} We perform photometric measurements using a circular aperture and an elongated 2D Gaussian. We compute circular aperture photometry with 
a) a diameter equal to the FWHM at each wavelength, 
b) a fixed aperture equal to the lowest resolution of our observations, i.e., $123$ mas ($2.5$ pc), and 
c) a fixed aperture encompassing the full extended emission of the Circinus, e.g., $640$ mas ($12.8$ pc). 
In addition, to optimize the extraction of the fluxes from the elongated emission, we performed photometric measurements using a 2D Gaussian profile. The 2D Gaussian profiles are fixed at the location of the peak pixel at each wavelength and have four free parameters: the x and y axes of the FWHM, the orientation, and the total amplitude of the peak of the 2D Gaussian profile. We computed a Markov Chain Monte Carlo (MCMC) approach using the No-U-Turn Sampler \citep[NUTS;][]{NUTS} method in the \textsc{python} code \textsc{pymc3} \citep{pymc3}. We set flat prior distributions within the ranges of $x=y=[0,3]\times$FWHM at a given wavelength, $\theta=[0,180)^{\circ}$ East of North, and $I_{0} = [0,1]$ (peak has been normalized to unity). We run the code using 5 chains with 5,000 steps and a 1,000 burn-in per chain, which provides 20,000 steps for the full MCMC code useful for data analysis. Figure \ref{fig:app_fig5} shows the best fits of the 2D Gaussian models per wavelength and the residuals. Table \ref{tab:app_table1} shows the photometric measurements of all the methods described above.

\begin{deluxetable}{lcccccc}[ht!]
\tablecaption{Photometric measurements. \label{tab:app_table1}}
\tablewidth{0pt}
\tablehead{\colhead{Filter}	& \colhead{Angular Aperture} & \colhead{Physical Aperture} &\colhead{Flux Density}	& \colhead{Contribution ext. emission} & \colhead{Flux density `North arc'} & \colhead{Contribution `North arc'}\\
	\colhead{} & \colhead{[mas]} & \colhead{[pc]} & \colhead{[mJy]} & \colhead{[\%]} & \colhead{[mJy]}  & \colhead{[\%]}\\
\colhead{(a)}  & \colhead{(b)} & \colhead{(c)} & \colhead{(d)} & \colhead{(e)} & \colhead{(f)} & \colhead{(g)}}
\startdata
\hline
F380M	& 	$93$             &	$1.9$	     & 	$250\pm10$		& - & - & -\\
	&	$123$             &	$2.5$	     &	$360\pm20$		& - & - & -\\
	&	$81\times162$, $-66^{\circ}$     &	$1.7\times3.3$	     &	$1050^{+60}_{-70}$ & $13^{+5}_{-6}$ & $14.1\pm0.7$ & $\sim1$\\ 
	&	$640$		     &	$12.8$	&	$1180\pm59$		& - & - & -\\
F430M	&	$105$		     &	$2.1$	&	$400\pm20$		& - & - & -\\
	&	$123$		     &	$2.5$	&	$510\pm30$		& - & - & -\\
	&	$84\times177$, $-68^{\circ}$	     &	$1.7\times3.6$	    &	$1670^{+90}_{-80}$	& $13^{+4}_{-5}$ & $18.9\pm0.9$ & $\sim1$\\ 
	&	$640$		     &	$12.8$	&	$1870\pm94$		& - & - & -\\
F480M	&	$123$		     &	$2.5$	&	$940\pm80$		& - & - & -\\
	&	$90\times183$, $-71^{\circ}$	    &	$1.8\times3.7$	&	$3030^{+150}_{-150}$ & $11^{+4}_{-4}$ & $21.0\pm1.7$ & $<1$\\ 
	&	$640$		  &	$12.8$	&	$3360\pm269$ & - & - & -
\enddata
\tablecomments{Columns, from left-hand to right-hand: 
(a) AMI/JWST filter, 
(b) angular circular aperture in units of mas and the x and y axes and PA of the 2D Gaussian profile,  
(c) physical circular aperture and the  x and y axes and PA of the 2D Gaussian profile, 
(d) flux density in units of mJy within the circular aperture and integrated under the full 2D Gaussian profile, 
(e) fractional contribution of all the extended emission after removal of the central 2D Gaussian profile,
(f) flux density in units of mJy of the `North arc' feature with a radius of $76$ mas ($1.5$ pc), and
(g) fractional contribution of the `North arc' feature within the central $10\times10$ pc$^{2}$ }
\end{deluxetable}

\begin{figure*}[ht!]
\includegraphics[width=\textwidth]{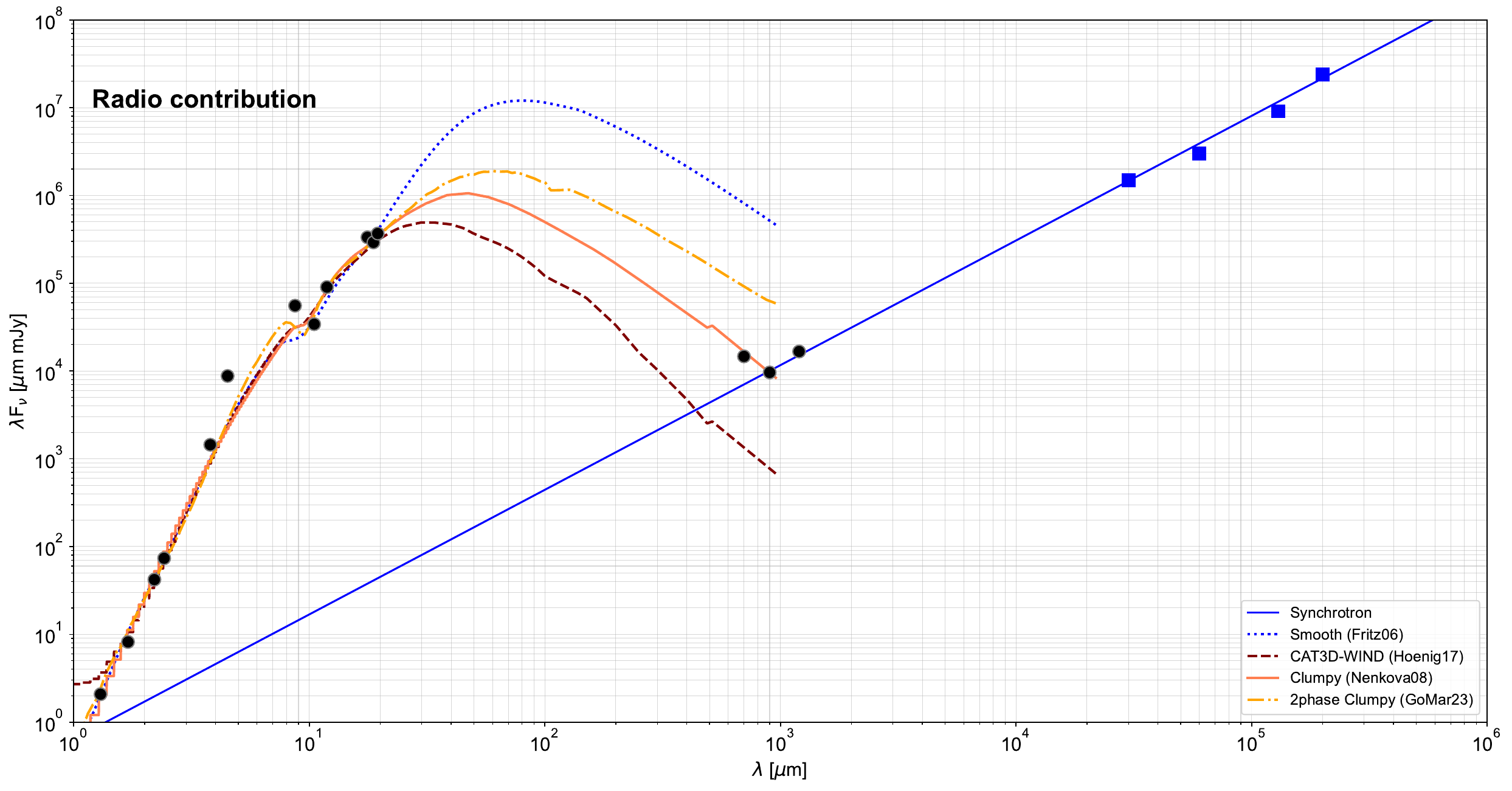}
\caption{\textbf{Radio contribution and torus models}. The Circinus SED (black circles) and the radio observations from ATCA (blue squares). The synchrotron emission passing through the ATCA data and the $1200\,\mu$m photometric points are shown (blue solid line).
 \label{fig:app_fig6}}
\end{figure*}

\begin{figure*}[ht!]
\includegraphics[width=\textwidth]{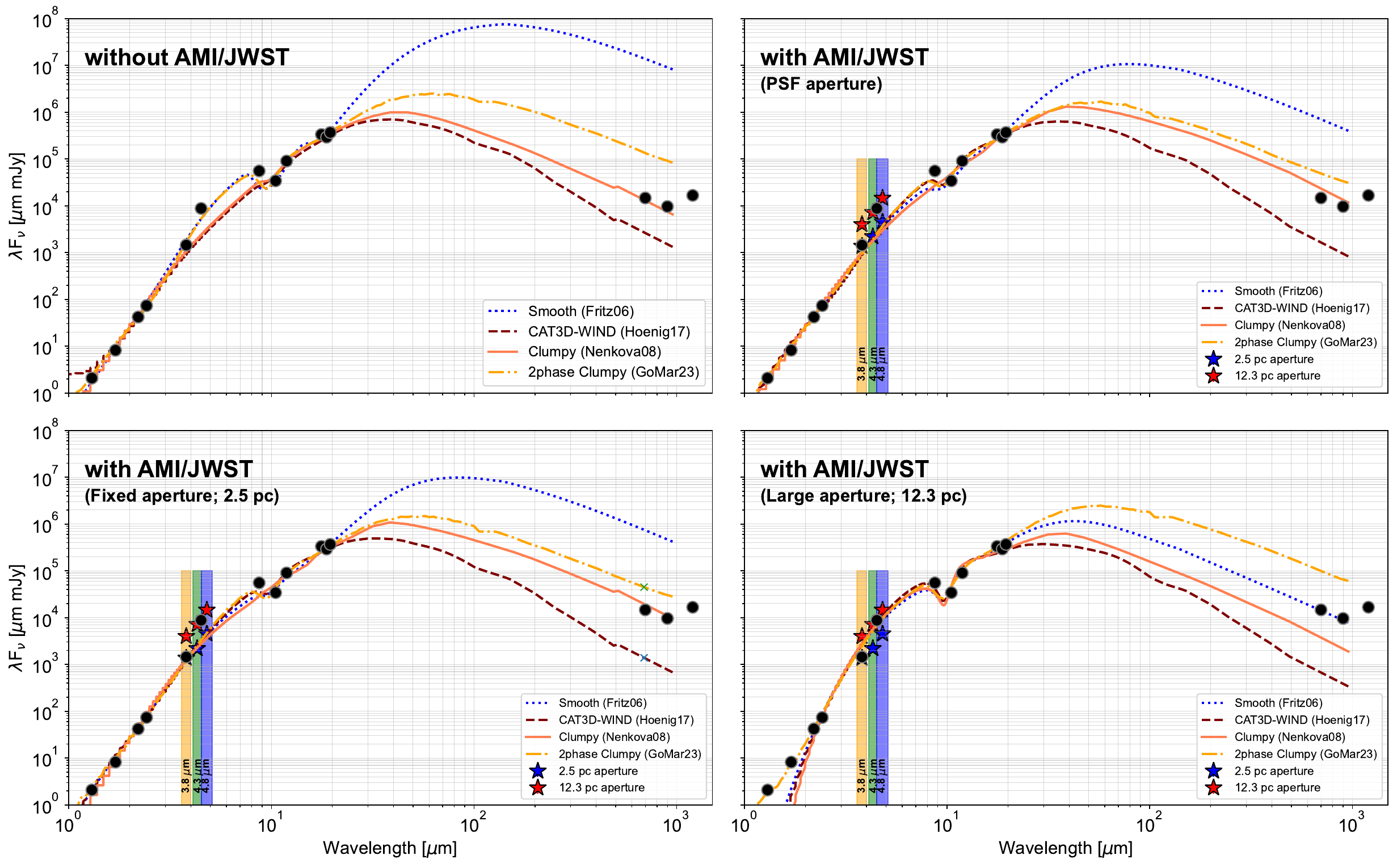}
\caption{\textbf{The $1-1000\,\mu$m SED of Circinus with best-fit torus models.} 
The SED without AMI/JWST data (top-left) and with the AMI/JWST photometric measurements fitting the PSF aperture (top-right), a fixed aperture of $1.5\,$pc (bottom left), a large aperture of $12.3\,$pc (bottom right) are shown. In each panel, we show the archival data (black circles), AMI/JWST photometry (stars), and the best-fit torus models as shown in the legend.
 \label{fig:app_fig7}}
\end{figure*}

\textbf{SED.} We took the $1-20\,\mu$m SED used by \citet{Stalevski2017}, and added the $3.8-4.8\,\mu$m AMI/JWST photometric points from our analysis (Table \ref{tab:app_table1}), and the photometric measurements of the central $123$ mas using the $700-1200\,\mu$m ALMA observations shown in Fig. \ref{fig:fig2}. The $1-1000\,\mu$m SED is shown in Figures \ref{fig:fig3}, \ref{fig:app_fig6}, and \ref{fig:app_fig7}. To estimate the relative contribution of non-thermal synchrotron emission at $700\,\mu$m, we use the radio observations from 3 to 20 cm observed by the Australia Telescope Compact Array (ATCA) \citep{Elmouttie1995}. We estimate that the non-thermal synchrotron emission at $700\,\mu$m contributes $<50$\%. Note that the ATCA data have a low angular resolution $\sim20"$, compared to the $\sim100$ mas resolution from the ALMA observations. Thus, this relative contribution is an overestimated upper-limit to the synchrotron emission at $700\,\mu$m.

\textbf{Torus models.} We took four torus models comprising several geometries with the main goal of distinguishing between a disk-like or wind-like structure \citep{GM2019,GB2022}. For all models, we fixed the inclination of the disk to be edge-on, $i=90^{\circ}$, and let the dust screen be a free parameter, E(B-V).

Smooth \citep{Fritz2006} torus model uses a torus-like geometry with a smooth dust distribution. The torus parameters are: $i$ is the viewing angle toward the torus, $\sigma$ is the half opening angle of the torus, $\gamma$ and $\beta$ are the exponents of the logarithmic azimuthal and radial density distributions, respectively, $Y=R_{o}/R_{i}$ is the ratio between the outer and inner radii of the torus, and $\tau_{V}$ is the edge-on optical depth at $0.55\,\mu$m.

\textsc{clumpy} \citep{Nenkova2008a,Nenkova2008b} torus model uses a clumpy distribution distributed in a torus-like structure. The free parameters are: $i$ is the viewing angle toward the torus, $N_{0}$ is the mean number of clouds radially across the equatorial plane, $\sigma$ is the half opening angle of the torus width measured from the equatorial plane, $Y=R_{o}/R_{i}$ is the ratio between the outer and inner radii of the torus, $q$ is the slope of the radial density distribution, and $\tau_{V}$ is the optical depth at $0.55\,\mu$m of individual clouds.

2-phase clumpy \citep{GM2023} torus model uses a torus geometry with high-density clumps and low-density, and a smooth interclump dust component. The free parameters are: $i$ is the viewing angle toward the torus, $\sigma$ is the half opening angle of the torus, $p$ and $q$ are the indices that set the dust density distributions along the radial and polar directions, respectively, $Y=R_{o}/R_{i}$ is the ratio between the outer and inner radii of the torus, $\tau_{V}$ is the average edge-on optical depth at $0.55\,\mu$m, and maximum dust grain sizes, $P_{\rm{size, max}}$. The extinction by dust grains was taken from \citep{Pei1992} and website \url{https://heasarc.gsfc.nasa.gov/xanadu/xspec/manual/node291.html}.

CAT3D-WIND \citep{Hoenig2017} torus model uses a clumpy disk and a dusty polar outflow. The free parameters are: $i$ is the viewing angle toward the torus, N$_{0}$ is the number of clouds along the equatorial plane, $a$ is the exponent of the radial distribution of clouds in the disk, $\theta$ is the half-opening angle of the dusty wind, $\sigma_{\theta}$ is the angular width of the hollow dusty wind cone, $a_{w}$ is the index of the dust cloud distribution power-law along the dusty wind, $h$ is the height of the inner edge of the torus, and $fwd$ is the wind-to-disk ratio.

We fit each of the torus models to the $1-1000\,\mu$m SED with and without the AMI/JWST photometric measurements (Fig. \ref{fig:app_fig7}). We use the same fitting routine described in \citet{GM2019}, which uses a $\chi^{2}$ minimization approach. The model parameters and $1\sigma$ uncertainties associated with the best-fit model are shown in Table \ref{tab:app_table2}. We estimate the $\chi^2$ of the models within the $1-1000\,\mu$m SED, $\chi^{2}_{ALL}$ and within the $1-20\,\mu$m SED, $\chi^{2}_{IR}$.

\begin{deluxetable}{lccccccccccc}[ht!]
\tablecaption{\textbf{Best-fit torus model parameters.}  \label{tab:app_table2}}
\tablewidth{0pt}
\tablehead{\colhead{\textsc{clumpy} (Nenkova08)} & \colhead{i}  & \colhead{N$_{0}$} & \colhead{$\sigma$} & \colhead{Y} & \colhead{q} & \colhead{$\tau_{V}$} & \colhead{E(B-V)} & \colhead{} & \colhead{} & \colhead{$\chi^{2}_{ALL}$} & \colhead{$\chi^{2}_{IR}$}}
\startdata
\hline
without AMI/JWST & $90^{\circ}$ & $4_{-1}^{+1}$ & $<70^{\circ}$ & $<100$ & $1.6_{-0.3}^{+0.3}$ & $24_{-5}^{+8}$ & $<0.11$ & & & $1.24$ & $1.23$\\
with AMI/JWST (psf) & $90^{\circ}$ & $4_{-1}^{+3}$ & $<70^{\circ}$ & $<95$ & $1.7_{-0.3}^{+0.4}$ & $30_{-3}^{+4}$ & $<0.19$ & & & $3.64$ & $3.21$ \\
with AMI/JWST (1.5pc) & $90^{\circ}$ & $4_{-1}^{+1}$ & $<70^{\circ}$ & $<100$ & $1.6_{-0.3}^{+0.3}$ & $20_{-2}^{+3}$ & $<0.12$ & & & $2.95$ & $2.93$ \\
with AMI/JWST (8pc) & $90^{\circ}$ & $<2$ & $60_{-45}^{+10}$$^{\circ}$ & $5_{-1}^{+2}$ & $<2.5$ & $59_{-23}^{+12}$ & $15.1_{-0.7}^{+0.6}$ & & & $6.54$ & $3.66$ \\
\hline
Smooth (Fritz06) & i & $\sigma$ & $\gamma$ & $\beta$ & Y & $\tau_{V}$ & E(B-V) & & & $\chi^{2}_{ALL}$ & $\chi^{2}_{IR}$\\
\hline
without AMI/JWST & $90^{\circ}$ & $33_{-5}^{+5}$ & $<0.01$ & $<-1$ & $56_{-3}^{+14}$ & $<10$ & $<0.45$  & & & $9\times10^{5}$ & $0.75$ \\
with AMI/JWST (psf) & $90^{\circ}$ & $23_{-3}^{+2}$ & $<0.01$ & $<-1$ & $47_{-9}^{+4}$ & $<10$ & $<0.13$  & & & $2\times10^{3}$ & $2.90$ \\
with AMI/JWST (1.5pc) & $90^{\circ}$ & $30_{-9}^{+3}$ & $<0.01$ & $<-1$ & $55_{-3}^{+23}$ & $<10$ & $<0.22$ & & & $2\times10^{3}$ & $2.28$ \\
with AMI/JWST (8pc) & $90^{\circ}$ & $23_{-3}^{+9}$ & $2.0_{-0.7}^{+0.3}$ & $<-1$ & $>10$ & $9.9_{-1.5}^{+0.1}$ & $7.7_{-0.6}^{+0.9}$ & & & $3.02$ & $3.01$ \\
\hline
2-phase clumpy (GoMar23) & i & $\sigma$ & p & q & Y & $\tau_{V}$ & $P_{\rm{size,max}}$ & E(B-V) & & $\chi^{2}_{ALL}$ & $\chi^{2}_{IR}$\\
\hline
without AMI/JWST & $90^{\circ}$ & $<11^{\circ}$ & $>1.3$ & $<1.5$ & $28_{-7}^{+2}$ & $6_{-1}^{+1}$ & $0.06_{-0.01}^{+0.01}$ & $<0.28$ & & $72.14$ & $0.61$ \\
with AMI/JWST (psf) & $90^{\circ}$ & $<80^{\circ}$ & $>1.5$ & $>1.0$ & $24_{-2}^{+4}$ & $5_{-1}^{+1}$ & $0.10_{-0.01}^{+0.05}$ & $1.1_{-0.1}^{+0.2}$ & & $8.70$ & $2.67$ \\
with AMI/JWST (1.5pc) & $90^{\circ}$ & $<80^{\circ}$ & $>1.3$ & $<1.5$ & $25_{-3}^{+4}$ & $5_{-1}^{+1}$ & $0.08_{-0.01}^{+0.01}$ & $0.69_{-0.20}^{+0.24}$ & & $6.54$ & $2.12$ \\
with AMI/JWST (8pc) & $90^{\circ}$ & $<80^{\circ}$ & $>1.4$ & $>1.4$ & $29_{-2}^{+1}$ & $6_{-1}^{+1}$ & $0.05_{-0.01}^{+0.01}$ & $<0.07$ & & $37.99$ & $1.40$ \\
\hline
CAT3D-WIND (Hoenig17) & i & N$_{0}$ & a & $\theta$ & $\sigma_{\theta}$ & a$_{w}$ & h & fwd & E(B-V) & $\chi^{2}_{ALL}$ & $\chi^{2}_{IR}$\\
\hline
without AMI/JWST & $90^{\circ}$ & $5.3_{-0.3}^{+0.8}$ & $-1.9_{-0.2}^{+0.3}$ & $45_{-7}^{+1}$$^{\circ}$ & $14_{-2}^{+2}$$^{\circ}$ & $<-2.5$ & $>0.1$ & $0.6_{-0.1}^{+0.1}$ & $2.5_{-0.4}^{+0.2}$ & $2.44$ & $1.79$ \\
with AMI/JWST (psf) & $90^{\circ}$ & $>5$ & $-1.6_{-0.2}^{+0.1}$ & $<45^{\circ}$ & $<15^{\circ}$ & $<-2.5$ & $>0.1$ & $0.5_{-0.6}^{+0.6}$ & $2.2_{-0.2}^{+0.2}$ & $3.69$ & $2.91$ \\
with AMI/JWST (1.5pc) & $90^{\circ}$ & $>5$ & $-1.8_{-0.2}^{+0.1}$ & $34_{-4}^{+11}$$^{\circ}$ & $<15^{\circ}$ & $<-2.5$ & $>0.1$ & $0.6_{-0.1}^{+0.6}$ & $2.3_{-0.2}^{+0.3}$ & $3.38$ & $2.53$ \\
with AMI/JWST (8pc) & $90^{\circ}$ & $7_{-2}^{+3}$ & $-2.6_{-0.1}^{+0.1}$ & $<30^{\circ}$ & $<15^{\circ}$ & $<-2.5$ & $>0.1$ & $<0.75$ & $10.8_{-1.0}^{+0.6}$ & $4.04$ & $3.13$ 
\enddata
\end{deluxetable}


\bibliography{references}{}

\begin{thebibliography}{}
\expandafter\ifx\csname natexlab\endcsname\relax\def\natexlab#1{#1}\fi
\providecommand{\url}[1]{\href{#1}{#1}}
\providecommand{\dodoi}[1]{doi:~\href{http://doi.org/#1}{\nolinkurl{#1}}}
\providecommand{\doeprint}[1]{\href{http://ascl.net/#1}{\nolinkurl{http://ascl.net/#1}}}
\providecommand{\doarXiv}[1]{\href{https://arxiv.org/abs/#1}{\nolinkurl{https://arxiv.org/abs/#1}}}

\bibitem[{A. {Alonso-Herrero} {et~al.}(2011){Alonso-Herrero}, {Ramos Almeida},
  {Mason}, {Asensio Ramos}, {Roche}, {Levenson}, {Elitzur}, {Packham},
  {Rodr{\'\i}guez Espinosa}, {Young}, {D{\'\i}az-Santos}, \&
  {P{\'e}rez-Garc{\'\i}a}}]{AH2011}
{Alonso-Herrero}, A., {Ramos Almeida}, C., {Mason}, R., {et~al.} 2011,
  \bibinfo{title}{{Torus and Active Galactic Nucleus Properties of Nearby
  Seyfert Galaxies: Results from Fitting Infrared Spectral Energy Distributions
  and Spectroscopy},} \apj, 736, 82,
  \dodoi{10.1088/0004-637X/736/2/8210.48550/arXiv.1105.2368}

\bibitem[{F. {Baron} {et~al.}(2010){Baron}, {Monnier}, \&
  {Kloppenborg}}]{squeeze2010}
{Baron}, F., {Monnier}, J.~D., \& {Kloppenborg}, B. 2010, \bibinfo{title}{{A
  novel image reconstruction software for optical/infrared interferometry},} in
  Society of Photo-Optical Instrumentation Engineers (SPIE) Conference Series,
  Vol. 7734, Optical and Infrared Interferometry II, ed. W.~C. {Danchi},
  F.~{Delplancke}, \& J.~K. {Rajagopal}, 77342I, \dodoi{10.1117/12.857364}

\bibitem[{G. {Cresci} {et~al.}(2015){Cresci}, {Marconi}, {Zibetti}, {Risaliti},
  {Carniani}, {Mannucci}, {Gallazzi}, {Maiolino}, {Balmaverde}, {Brusa},
  {Capetti}, {Cicone}, {Feruglio}, {Bland-Hawthorn}, {Nagao}, {Oliva},
  {Salvato}, {Sani}, {Tozzi}, {Urrutia}, \& {Venturi}}]{Cresci2015}
{Cresci}, G., {Marconi}, A., {Zibetti}, S., {et~al.} 2015, \bibinfo{title}{{The
  MAGNUM survey: positive feedback in the nuclear region of NGC 5643 suggested
  by MUSE},} \aap, 582, A63,
  \dodoi{10.1051/0004-6361/20152658110.48550/arXiv.1508.04464}

\bibitem[{G.~P. {Donnelly} {et~al.}(2025){Donnelly}, {Lai}, {Armus},
  {D{\'\i}az-Santos}, {Larson}, {Barcos-Mu{\~n}oz}, {Bianchin}, {Bohn},
  {B{\"o}ker}, {Buiten}, {Charmandaris}, {Evans}, {Howell}, {Inami}, {Kakkad},
  {Lenki{\'c}}, {Linden}, {Lofaro}, {Malkan}, {Medling}, {Privon}, {Ricci},
  {Smith}, {Song}, {Stierwalt}, {van der Werf}, \& {U}}]{Donnelly25}
{Donnelly}, G.~P., {Lai}, T. S.~Y., {Armus}, L., {et~al.} 2025,
  \bibinfo{title}{{A Spectroscopically Calibrated Prescription for Extracting
  PAH Flux from JWST MIRI Imaging},} arXiv e-prints, arXiv:2501.19397,
  \dodoi{10.48550/arXiv.2501.19397}

\bibitem[{R. {Doyon} {et~al.}(2012){Doyon}, {Hutchings}, {Beaulieu}, {Albert},
  {Lafreni{\`e}re}, {Willott}, {Touahri}, {Rowlands}, {Maszkiewicz},
  {Fullerton}, {Volk}, {Martel}, {Chayer}, {Sivaramakrishnan}, {Abraham},
  {Ferrarese}, {Jayawardhana}, {Johnstone}, {Meyer}, {Pipher}, \&
  {Sawicki}}]{NIRISS2012}
{Doyon}, R., {Hutchings}, J.~B., {Beaulieu}, M., {et~al.} 2012,
  \bibinfo{title}{{The JWST Fine Guidance Sensor (FGS) and Near-Infrared Imager
  and Slitless Spectrograph (NIRISS)},} in Society of Photo-Optical
  Instrumentation Engineers (SPIE) Conference Series, Vol. 8442, Space
  Telescopes and Instrumentation 2012: Optical, Infrared, and Millimeter Wave,
  ed. M.~C. {Clampin}, G.~G. {Fazio}, H.~A. {MacEwen}, \& J.~{Oschmann},
  Jacobus~M., 84422R, \dodoi{10.1117/12.926578}

\bibitem[{Y. {Dubois} {et~al.}(2016){Dubois}, {Peirani}, {Pichon}, {Devriendt},
  {Gavazzi}, {Welker}, \& {Volonteri}}]{Dubois2016}
{Dubois}, Y., {Peirani}, S., {Pichon}, C., {et~al.} 2016, \bibinfo{title}{{The
  HORIZON-AGN simulation: morphological diversity of galaxies promoted by AGN
  feedback},} \mnras, 463, 3948, \dodoi{10.1093/mnras/stw2265}

\bibitem[{E. {Elmouttie} {et~al.}(1995){Elmouttie}, {Haynes}, {Jones}, {Ehle},
  {Beck}, \& {Wielebinski}}]{Elmouttie1995}
{Elmouttie}, E., {Haynes}, R.~F., {Jones}, K.~L., {et~al.} 1995,
  \bibinfo{title}{{The polarized radio lobes of the Circinus galaxy},} \mnras,
  275, L53, \dodoi{10.1093/mnras/275.1.L53}

\bibitem[{M. {Elmouttie} {et~al.}(1998){Elmouttie}, {Haynes}, {Jones},
  {Sadler}, \& {Ehle}}]{Elmouttie1998}
{Elmouttie}, M., {Haynes}, R.~F., {Jones}, K.~L., {Sadler}, E.~M., \& {Ehle},
  M. 1998, \bibinfo{title}{{Radio continuum evidence for nuclear outflow in the
  Circinus galaxy},} \mnras, 297, 1202,
  \dodoi{10.1046/j.1365-8711.1998.01592.x}

\bibitem[{R.~T. {Emmering} {et~al.}(1992){Emmering}, {Blandford}, \&
  {Shlosman}}]{Emmering1992}
{Emmering}, R.~T., {Blandford}, R.~D., \& {Shlosman}, I. 1992,
  \bibinfo{title}{{Magnetic Acceleration of Broad Emission-Line Clouds in
  Active Galactic Nuclei},} \apj, 385, 460, \dodoi{10.1086/170955}

\bibitem[{J. {Fritz} {et~al.}(2006){Fritz}, {Franceschini}, \&
  {Hatziminaoglou}}]{Fritz2006}
{Fritz}, J., {Franceschini}, A., \& {Hatziminaoglou}, E. 2006,
  \bibinfo{title}{{Revisiting the infrared spectra of active galactic nuclei
  with a new torus emission model},} \mnras, 366, 767,
  \dodoi{10.1111/j.1365-2966.2006.09866.x}

\bibitem[{J.~F. {Gallimore} {et~al.}(2024){Gallimore}, {Impellizzeri},
  {Aghelpasand}, {Gao}, {Hostetter}, \& {Lankhaar}}]{Gallimore2024}
{Gallimore}, J.~F., {Impellizzeri}, C.~M.~V., {Aghelpasand}, S., {et~al.} 2024,
  \bibinfo{title}{{The Discovery of Polarized Water Vapor Megamaser Emission in
  a Molecular Accretion Disk},} \apjl, 975, L9,
  \dodoi{10.3847/2041-8213/ad864f}

\bibitem[{V. {G{\'a}mez Rosas} {et~al.}(2022){G{\'a}mez Rosas}, {Isbell},
  {Jaffe}, {Petrov}, {Leftley}, {Hofmann}, {Millour}, {Burtscher},
  {Meisenheimer}, {Meilland}, {Waters}, {Lopez}, {Lagarde}, {Weigelt}, {Berio},
  {Allouche}, {Robbe-Dubois}, {Cruzal{\`e}bes}, {Bettonvil}, {Henning},
  {Augereau}, {Antonelli}, {Beckmann}, {van Boekel}, {Bendjoya}, {Danchi},
  {Dominik}, {Drevon}, {Gallimore}, {Graser}, {Heininger}, {Hocd{\'e}},
  {Hogerheijde}, {Hron}, {Impellizzeri}, {Klarmann}, {Kokoulina}, {Labadie},
  {Lehmitz}, {Matter}, {Paladini}, {Pantin}, {Pott}, {Schertl}, {Soulain},
  {Stee}, {Tristram}, {Varga}, {Woillez}, {Wolf}, {Yoffe}, \& {Zins}}]{GR2022}
{G{\'a}mez Rosas}, V., {Isbell}, J.~W., {Jaffe}, W., {et~al.} 2022,
  \bibinfo{title}{{Thermal imaging of dust hiding the black hole in NGC 1068},}
  \nat, 602, 403, \dodoi{10.1038/s41586-021-04311-710.48550/arXiv.2112.13694}

\bibitem[{I. {Garc{\'\i}a-Bernete} {et~al.}(2022){Garc{\'\i}a-Bernete},
  {Gonz{\'a}lez-Mart{\'\i}n}, {Ramos Almeida}, {Alonso-Herrero},
  {Mart{\'\i}nez-Paredes}, {Ward}, {Roche}, {Acosta-Pulido},
  {L{\'o}pez-Rodr{\'\i}guez}, {Rigopoulou}, \& {Esparza-Arredondo}}]{GB2022}
{Garc{\'\i}a-Bernete}, I., {Gonz{\'a}lez-Mart{\'\i}n}, O., {Ramos Almeida}, C.,
  {et~al.} 2022, \bibinfo{title}{{Torus and polar dust dependence on active
  galactic nucleus properties},} \aap, 667, A140,
  \dodoi{10.1051/0004-6361/202244230}

\bibitem[{I. {Garc{\'\i}a-Bernete}
  {et~al.}(2024{\natexlab{a}}){Garc{\'\i}a-Bernete}, {Alonso-Herrero},
  {Rigopoulou}, {Pereira-Santaella}, {Shimizu}, {Davies}, {Donnan}, {Roche},
  {Gonz{\'a}lez-Mart{\'\i}n}, {Ramos Almeida}, {Bellocchi}, {Boorman},
  {Combes}, {Efstathiou}, {Esparza-Arredondo}, {Garc{\'\i}a-Burillo},
  {Gonz{\'a}lez-Alfonso}, {Hicks}, {H{\"o}nig}, {Labiano}, {Levenson},
  {L{\'o}pez-Rodr{\'\i}guez}, {Ricci}, {Packham}, {Rouan}, {Stalevski}, \&
  {Ward}}]{Bernete24b}
{Garc{\'\i}a-Bernete}, I., {Alonso-Herrero}, A., {Rigopoulou}, D., {et~al.}
  2024{\natexlab{a}}, \bibinfo{title}{{The Galaxy Activity, Torus, and Outflow
  Survey (GATOS). III. Revealing the inner icy structure in local active
  galactic nuclei},} \aap, 681, L7, \dodoi{10.1051/0004-6361/202348266}

\bibitem[{I. {Garc{\'\i}a-Bernete}
  {et~al.}(2024{\natexlab{b}}){Garc{\'\i}a-Bernete}, {Rigopoulou}, {Donnan},
  {Alonso-Herrero}, {Pereira-Santaella}, {Shimizu}, {Davies}, {Roche},
  {Garc{\'\i}a-Burillo}, {Labiano}, {Hermosa Mu{\~n}oz}, {Zhang}, {Audibert},
  {Bellocchi}, {Bunker}, {Combes}, {Delaney}, {Esparza-Arredondo}, {Gandhi},
  {Gonz{\'a}lez-Mart{\'\i}n}, {H{\"o}nig}, {Imanishi}, {Hicks}, {Fuller},
  {Leist}, {Levenson}, {Lopez-Rodriguez}, {Packham}, {Ramos Almeida}, {Ricci},
  {Stalevski}, {Villar Mart{\'\i}n}, \& {Ward}}]{Bernete24c}
{Garc{\'\i}a-Bernete}, I., {Rigopoulou}, D., {Donnan}, F.~R., {et~al.}
  2024{\natexlab{b}}, \bibinfo{title}{{The Galaxy Activity, Torus, and Outflow
  Survey (GATOS): V. Unveiling PAH survival and resilience in the circumnuclear
  regions of AGNs with JWST},} \aap, 691, A162,
  \dodoi{10.1051/0004-6361/202450086}

\bibitem[{S. {Garc{\'\i}a-Burillo} {et~al.}(2024){Garc{\'\i}a-Burillo},
  {Hicks}, {Alonso-Herrero}, {Pereira-Santaella}, {Usero}, {Querejeta},
  {Gonz{\'a}lez-Mart{\'\i}n}, {Delaney}, {Ramos Almeida}, {Combes},
  {Angl{\'e}s-Alc{\'a}zar}, {Audibert}, {Bellocchi}, {Davies}, {Davis},
  {Elford}, {Garc{\'\i}a-Bernete}, {H{\"o}nig}, {Labiano}, {Leist}, {Levenson},
  {L{\'o}pez-Rodr{\'\i}guez}, {Mercedes-Feliz}, {Packham}, {Ricci}, {Rosario},
  {Shimizu}, {Stalevski}, \& {Zhang}}]{GB2024}
{Garc{\'\i}a-Burillo}, S., {Hicks}, E.~K.~S., {Alonso-Herrero}, A., {et~al.}
  2024, \bibinfo{title}{{Deciphering the imprint of active galactic nucleus
  feedback in Seyfert galaxies: Nuclear-scale molecular gas deficits},} \aap,
  689, A347, \dodoi{10.1051/0004-6361/202450268}

\bibitem[{E. {Gonz{\'a}lez-Alfonso} {et~al.}(2024){Gonz{\'a}lez-Alfonso},
  {Garc{\'\i}a-Bernete}, {Pereira-Santaella}, {Neufeld}, {Fischer}, \&
  {Donnan}}]{Gonzalez-Alfonso24}
{Gonz{\'a}lez-Alfonso}, E., {Garc{\'\i}a-Bernete}, I., {Pereira-Santaella}, M.,
  {et~al.} 2024, \bibinfo{title}{{JWST detection of extremely excited
  outflowing CO and H$_{2}$O in VV 114 E SW: A possible rapidly accreting
  IMBH},} \aap, 682, A182, \dodoi{10.1051/0004-6361/202348469}

\bibitem[{O. {Gonz{\'a}lez-Mart{\'\i}n}
  {et~al.}(2019){Gonz{\'a}lez-Mart{\'\i}n}, {Masegosa}, {Garc{\'\i}a-Bernete},
  {Ramos Almeida}, {Rodr{\'\i}guez-Espinosa}, {M{\'a}rquez},
  {Esparza-Arredondo}, {Osorio-Clavijo}, {Mart{\'\i}nez-Paredes},
  {Victoria-Ceballos}, {Pasetto}, \& {Dultzin}}]{GM2019}
{Gonz{\'a}lez-Mart{\'\i}n}, O., {Masegosa}, J., {Garc{\'\i}a-Bernete}, I.,
  {et~al.} 2019, \bibinfo{title}{{Exploring the Mid-infrared SEDs of Six AGN
  Dusty Torus Models. I. Synthetic Spectra},} \apj, 884, 10,
  \dodoi{10.3847/1538-4357/ab3e6b}

\bibitem[{O. {Gonz{\'a}lez-Mart{\'\i}n}
  {et~al.}(2023){Gonz{\'a}lez-Mart{\'\i}n}, {Ramos Almeida}, {Fritz},
  {Alonso-Herrero}, {H{\"o}nig}, {Roche}, {Esparza-Arredondo},
  {Garc{\'\i}a-Bernete}, {Garc{\'\i}a-Burillo}, {Osorio-Clavijo},
  {Reyes-Amador}, {Stalevski}, \& {Victoria-Ceballos}}]{GM2023}
{Gonz{\'a}lez-Mart{\'\i}n}, O., {Ramos Almeida}, C., {Fritz}, J., {et~al.}
  2023, \bibinfo{title}{{The role of grain size in active galactic nuclei torus
  dust models},} \aap, 676, A73, \dodoi{10.1051/0004-6361/202345858}

\bibitem[{L.~J. {Greenhill} {et~al.}(2003){Greenhill}, {Booth}, {Ellingsen},
  {Herrnstein}, {Jauncey}, {McCulloch}, {Moran}, {Norris}, {Reynolds}, \&
  {Tzioumis}}]{Greenhill2003}
{Greenhill}, L.~J., {Booth}, R.~S., {Ellingsen}, S.~P., {et~al.} 2003,
  \bibinfo{title}{{A Warped Accretion Disk and Wide-Angle Outflow in the Inner
  Parsec of the Circinus Galaxy},} \apj, 590, 162, \dodoi{10.1086/374862}

\bibitem[{M.~D. Homan \& A. Gelman(2014)Homan \& Gelman}]{NUTS}
Homan, M.~D., \& Gelman, A. 2014, \bibinfo{title}{The No-U-turn sampler:
  adaptively setting path lengths in Hamiltonian Monte Carlo,} J. Mach. Learn.
  Res., 15, 1593–1623

\bibitem[{S.~F. {H{\"o}nig}(2019){H{\"o}nig}}]{Honig2019}
{H{\"o}nig}, S.~F. 2019, \bibinfo{title}{{Redefining the Torus: A Unifying View
  of AGNs in the Infrared and Submillimeter},} \apj, 884, 171,
  \dodoi{10.3847/1538-4357/ab459110.48550/arXiv.1909.08639}

\bibitem[{S.~F. {H{\"o}nig} \& M. {Kishimoto}(2017){H{\"o}nig} \&
  {Kishimoto}}]{Hoenig2017}
{H{\"o}nig}, S.~F., \& {Kishimoto}, M. 2017, \bibinfo{title}{{Dusty Winds in
  Active Galactic Nuclei: Reconciling Observations with Models},} \apjl, 838,
  L20, \dodoi{10.3847/2041-8213/aa6838}

\bibitem[{P.~F. {Hopkins} {et~al.}(2016){Hopkins}, {Torrey},
  {Faucher-Gigu{\`e}re}, {Quataert}, \& {Murray}}]{Hopkins2016}
{Hopkins}, P.~F., {Torrey}, P., {Faucher-Gigu{\`e}re}, C.-A., {Quataert}, E.,
  \& {Murray}, N. 2016, \bibinfo{title}{{Stellar and quasar feedback in
  concert: effects on AGN accretion, obscuration, and outflows},} \mnras, 458,
  816, \dodoi{10.1093/mnras/stw28910.48550/arXiv.1504.05209}

\bibitem[{J.~W. {Isbell} {et~al.}(2022){Isbell}, {Meisenheimer}, {Pott},
  {Stalevski}, {Tristram}, {Sanchez-Bermudez}, {Hofmann}, {G{\'a}mez Rosas},
  {Jaffe}, {Burtscher}, {Leftley}, {Petrov}, {Lopez}, {Henning}, {Weigelt},
  {Allouche}, {Berio}, {Bettonvil}, {Cruzalebes}, {Dominik}, {Heininger},
  {Hogerheijde}, {Lagarde}, {Lehmitz}, {Matter}, {Meilland}, {Millour},
  {Robbe-Dubois}, {Schertl}, {van Boekel}, {Varga}, \& {Woillez}}]{Isbell2022}
{Isbell}, J.~W., {Meisenheimer}, K., {Pott}, J.~U., {et~al.} 2022,
  \bibinfo{title}{{The dusty heart of Circinus. I. Imaging the circumnuclear
  dust in N-band},} \aap, 663, A35, \dodoi{10.1051/0004-6361/202243271}

\bibitem[{J.~W. {Isbell} {et~al.}(2023){Isbell}, {Pott}, {Meisenheimer},
  {Stalevski}, {Tristram}, {Leftley}, {Asmus}, {Weigelt}, {G{\'a}mez Rosas},
  {Petrov}, {Jaffe}, {Hofmann}, {Henning}, \& {Lopez}}]{Isbell2023}
{Isbell}, J.~W., {Pott}, J.~U., {Meisenheimer}, K., {et~al.} 2023,
  \bibinfo{title}{{The dusty heart of Circinus. II. Scrutinizing the LM-band
  dust morphology using MATISSE},} \aap, 678, A136,
  \dodoi{10.1051/0004-6361/202347307}

\bibitem[{T. {Izumi} {et~al.}(2023){Izumi}, {Wada}, {Imanishi}, {Nakanishi},
  {Kohno}, {Kudoh}, {Kawamuro}, {Baba}, {Matsumoto}, {Fujita}, \&
  {Tristram}}]{Izumi2023}
{Izumi}, T., {Wada}, K., {Imanishi}, M., {et~al.} 2023,
  \bibinfo{title}{{Supermassive black hole feeding and feedback observed on
  subparsec scales},} Science, 382, 554, \dodoi{10.1126/science.adf0569}

\bibitem[{R.~C. {Jennison} \& V. {Latham}(1959){Jennison} \&
  {Latham}}]{Jennison1959}
{Jennison}, R.~C., \& {Latham}, V. 1959, \bibinfo{title}{{The brightness
  distribution within the radio sources Cygnus A (19N4A) and Cassiopeia A
  (23N5A)},} \mnras, 119, 174, \dodoi{10.1093/mnras/119.2.174}

\bibitem[{M. {Kishimoto} {et~al.}(2009){Kishimoto}, {H{\"o}nig}, {Antonucci},
  {Kotani}, {Barvainis}, {Tristram}, \& {Weigelt}}]{Kishimoto2009}
{Kishimoto}, M., {H{\"o}nig}, S.~F., {Antonucci}, R., {et~al.} 2009,
  \bibinfo{title}{{Exploring the inner region of type 1 AGNs with the Keck
  interferometer},} \aap, 507, L57,
  \dodoi{10.1051/0004-6361/20091351210.48550/arXiv.0911.0666}

\bibitem[{M. {Kishimoto} {et~al.}(2022){Kishimoto}, {Anderson}, {ten
  Brummelaar}, {Farrington}, {Antonucci}, {H{\"o}nig}, {Millour}, {Tristram},
  {Weigelt}, {Sturmann}, {Sturmann}, {Schaefer}, \& {Scott}}]{Kishimoto2022}
{Kishimoto}, M., {Anderson}, M., {ten Brummelaar}, T., {et~al.} 2022,
  \bibinfo{title}{{The Dust Sublimation Region of the Type 1 AGN NGC 4151 at a
  Hundred Microarcsecond Scale as Resolved by the CHARA Array Interferometer},}
  \apj, 940, 28, \dodoi{10.3847/1538-4357/ac91c410.48550/arXiv.2209.06061}

\bibitem[{R.~M. {Lau} {et~al.}(2024){Lau}, {Hankins}, {Sanchez-Bermudez},
  {Thatte}, {Soulain}, {Cooper}, {Sivaramakrishnan}, {Corcoran}, {Greenbaum},
  {Gull}, {Han}, {Jones}, {Madura}, {Moffat}, {Morris}, {Onaka}, {Russell},
  {Richardson}, {Smith}, {Tuthill}, {Volk}, {Weigelt}, \& {Williams}}]{Lau2024}
{Lau}, R.~M., {Hankins}, M.~J., {Sanchez-Bermudez}, J., {et~al.} 2024,
  \bibinfo{title}{{A First Look with JWST Aperture Masking Interferometry:
  Resolving Circumstellar Dust around the Wolf{\textendash}Rayet Binary WR 137
  beyond the Rayleigh Limit},} \apj, 963, 127, \dodoi{10.3847/1538-4357/ad192c}

\bibitem[{E. {Lopez-Rodriguez} {et~al.}(2023){Lopez-Rodriguez}, {Kishimoto},
  {Antonucci}, {Begelman}, {Globus}, \& {Blandford}}]{ELR2023}
{Lopez-Rodriguez}, E., {Kishimoto}, M., {Antonucci}, R., {et~al.} 2023,
  \bibinfo{title}{{On the Origin of Radio-loudness in Active Galactic Nuclei
  Using Far-infrared Polarimetric Observations},} \apj, 951, 31,
  \dodoi{10.3847/1538-4357/accb96}

\bibitem[{E. {Lopez-Rodriguez} {et~al.}(2015){Lopez-Rodriguez}, {Packham},
  {Jones}, {Nikutta}, {McMaster}, {Mason}, {Elvis}, {Shenoy}, {Alonso-Herrero},
  {Ram{\'\i}rez}, {Gonz{\'a}lez Mart{\'\i}n}, {H{\"o}nig}, {Levenson}, {Ramos
  Almeida}, \& {Perlman}}]{ELR2015}
{Lopez-Rodriguez}, E., {Packham}, C., {Jones}, T.~J., {et~al.} 2015,
  \bibinfo{title}{{Near-infrared polarimetric adaptive optics observations of
  NGC 1068: a torus created by a hydromagnetic outflow wind},} \mnras, 452,
  1902, \dodoi{10.1093/mnras/stv1410}

\bibitem[{E. {Lopez-Rodriguez} {et~al.}(2018){Lopez-Rodriguez}, {Fuller},
  {Alonso-Herrero}, {Efstathiou}, {Ichikawa}, {Levenson}, {Packham},
  {Radomski}, {Ramos Almeida}, {Benford}, {Berthoud}, {Hamilton}, {Harper},
  {Kov{\'a}vcs}, {Santos}, {Staguhn}, \& {Herter}}]{ELR2018}
{Lopez-Rodriguez}, E., {Fuller}, L., {Alonso-Herrero}, A., {et~al.} 2018,
  \bibinfo{title}{{The Emission and Distribution of Dust of the Torus of NGC
  1068},} \apj, 859, 99, \dodoi{10.3847/1538-4357/aabd7b}

\bibitem[{A.~F.~M. {Moorwood} {et~al.}(1996){Moorwood}, {Lutz}, {Oliva},
  {Marconi}, {Netzer}, {Genzel}, {Sturm}, \& {de Graauw}}]{Moorwood1996}
{Moorwood}, A.~F.~M., {Lutz}, D., {Oliva}, E., {et~al.} 1996,
  \bibinfo{title}{{2.5-45{\ensuremath{\mu}}m SWS spectroscopy of the Circinus
  Galaxy.},} \aap, 315, L109

\bibitem[{R. {Mor} {et~al.}(2009){Mor}, {Netzer}, \& {Elitzur}}]{Mor2009}
{Mor}, R., {Netzer}, H., \& {Elitzur}, M. 2009, \bibinfo{title}{{Dusty
  Structure Around Type-I Active Galactic Nuclei: Clumpy Torus Narrow-line
  Region and Near-nucleus Hot Dust},} \apj, 705, 298,
  \dodoi{10.1088/0004-637X/705/1/29810.48550/arXiv.0907.1654}

\bibitem[{R. {Morganti}(2017){Morganti}}]{Morganti2017}
{Morganti}, R. 2017, \bibinfo{title}{{The many routes to AGN feedback},}
  Frontiers in Astronomy and Space Sciences, 4, 42,
  \dodoi{10.3389/fspas.2017.00042}

\bibitem[{M. {Nenkova} {et~al.}(2008{\natexlab{a}}){Nenkova}, {Sirocky},
  {Ivezi{\'c}}, \& {Elitzur}}]{Nenkova2008a}
{Nenkova}, M., {Sirocky}, M.~M., {Ivezi{\'c}}, {\v{Z}}., \& {Elitzur}, M.
  2008{\natexlab{a}}, \bibinfo{title}{{AGN Dusty Tori. I. Handling of Clumpy
  Media},} \apj, 685, 147, \dodoi{10.1086/590482}

\bibitem[{M. {Nenkova} {et~al.}(2008{\natexlab{b}}){Nenkova}, {Sirocky},
  {Nikutta}, {Ivezi{\'c}}, \& {Elitzur}}]{Nenkova2008b}
{Nenkova}, M., {Sirocky}, M.~M., {Nikutta}, R., {Ivezi{\'c}}, {\v{Z}}., \&
  {Elitzur}, M. 2008{\natexlab{b}}, \bibinfo{title}{{AGN Dusty Tori. II.
  Observational Implications of Clumpiness},} \apj, 685, 160,
  \dodoi{10.1086/590483}

\bibitem[{R. {Nikutta} {et~al.}(2021{\natexlab{a}}){Nikutta},
  {Lopez-Rodriguez}, {Ichikawa}, {Levenson}, {Packham}, {H{\"o}nig}, \&
  {Alonso-Herrero}}]{HypercatI}
{Nikutta}, R., {Lopez-Rodriguez}, E., {Ichikawa}, K., {et~al.}
  2021{\natexlab{a}}, \bibinfo{title}{{Hypercubes of AGN Tori (HYPERCAT). I.
  Models and Image Morphology},} \apj, 919, 136,
  \dodoi{10.3847/1538-4357/ac06a6}

\bibitem[{R. {Nikutta} {et~al.}(2021{\natexlab{b}}){Nikutta},
  {Lopez-Rodriguez}, {Ichikawa}, {Levenson}, {Packham}, {H{\"o}nig}, \&
  {Alonso-Herrero}}]{HypercatII}
{Nikutta}, R., {Lopez-Rodriguez}, E., {Ichikawa}, K., {et~al.}
  2021{\natexlab{b}}, \bibinfo{title}{{Hypercubes of AGN Tori (HYPERCAT). II.
  Resolving the Torus with Extremely Large Telescopes},} \apj, 923, 127,
  \dodoi{10.3847/1538-4357/ac2949}

\bibitem[{E. {Oliva} {et~al.}(1999){Oliva}, {Marconi}, \&
  {Moorwood}}]{Oliva1999}
{Oliva}, E., {Marconi}, A., \& {Moorwood}, A.~F.~M. 1999,
  \bibinfo{title}{{Metal abundances and excitation of extranuclear clouds in
  the Circinus galaxy. A new method for deriving abundances of AGN narrow line
  clouds},} \aap, 342, 87, \dodoi{10.48550/arXiv.astro-ph/9811177}

\bibitem[{T.~A. {Pauls} {et~al.}(2005){Pauls}, {Young}, {Cotton}, \&
  {Monnier}}]{Pauls_2005}
{Pauls}, T.~A., {Young}, J.~S., {Cotton}, W.~D., \& {Monnier}, J.~D. 2005,
  \bibinfo{title}{{A Data Exchange Standard for Optical (Visible/IR)
  Interferometry},} \pasp, 117, 1255, \dodoi{10.1086/444523}

\bibitem[{Y.~C. {Pei}(1992){Pei}}]{Pei1992}
{Pei}, Y.~C. 1992, \bibinfo{title}{{Interstellar Dust from the Milky Way to the
  Magellanic Clouds},} \apj, 395, 130, \dodoi{10.1086/171637}

\bibitem[{M. {Pereira-Santaella} {et~al.}(2024){Pereira-Santaella},
  {Gonz{\'a}lez-Alfonso}, {Garc{\'\i}a-Bernete}, {Garc{\'\i}a-Burillo}, \&
  {Rigopoulou}}]{Pereira-Santaella24}
{Pereira-Santaella}, M., {Gonz{\'a}lez-Alfonso}, E., {Garc{\'\i}a-Bernete}, I.,
  {Garc{\'\i}a-Burillo}, S., \& {Rigopoulou}, D. 2024, \bibinfo{title}{{The
  CO-to-H$_{2}$ conversion factor of molecular outflows. Rovibrational CO
  emission in NGC 3256-S resolved by JWST/NIRSpec},} \aap, 681, A117,
  \dodoi{10.1051/0004-6361/202347942}

\bibitem[{E.~A. {Pier} \& J.~H. {Krolik}(1993){Pier} \& {Krolik}}]{PK1993}
{Pier}, E.~A., \& {Krolik}, J.~H. 1993, \bibinfo{title}{{Infrared Spectra of
  Obscuring Dust Tori around Active Galactic Nuclei. II. Comparison with
  Observations},} \apj, 418, 673, \dodoi{10.1086/173427}

\bibitem[{M.~A. {Prieto} {et~al.}(2010){Prieto}, {Reunanen}, {Tristram},
  {Neumayer}, {Fernandez-Ontiveros}, {Orienti}, \& {Meisenheimer}}]{Prieto2010}
{Prieto}, M.~A., {Reunanen}, J., {Tristram}, K.~R.~W., {et~al.} 2010,
  \bibinfo{title}{{The spectral energy distribution of the central parsecs of
  the nearest AGN},} \mnras, 402, 724, \dodoi{10.1111/j.1365-2966.2009.15897.x}

\bibitem[{M.~A. {Prieto} {et~al.}(2004){Prieto}, {Meisenheimer}, {Marco},
  {Reunanen}, {Contini}, {Clenet}, {Davies}, {Gratadour}, {Henning}, {Klaas},
  {Kotilainen}, {Leinert}, {Lutz}, {Rouan}, \& {Thatte}}]{Prieto2004}
{Prieto}, M.~A., {Meisenheimer}, K., {Marco}, O., {et~al.} 2004,
  \bibinfo{title}{{Unveiling the Central Parsec Region of an Active Galactic
  Nucleus: The Circinus Nucleus in the Near-Infrared with the Very Large
  Telescope},} \apj, 614, 135, \dodoi{10.1086/423422}

\bibitem[{A.~C.~S. {Readhead} {et~al.}(1988){Readhead}, {Nakajima}, {Pearson},
  {Neugebauer}, {Oke}, \& {Sargent}}]{Readhead1988}
{Readhead}, A.~C.~S., {Nakajima}, T.~S., {Pearson}, T.~J., {et~al.} 1988,
  \bibinfo{title}{{Diffraction-Limited Imaging with Ground-Based Optical
  Telescopes},} \aj, 95, 1278, \dodoi{10.1086/114724}

\bibitem[{J. {Salvatier} {et~al.}(2016){Salvatier}, {Wieckia}, \&
  {Fonnesbeck}}]{pymc3}
{Salvatier}, J., {Wieckia}, T.~V., \& {Fonnesbeck}, C. 2016, {PyMC3: Python
  probabilistic programming framework},, Astrophysics Source Code Library,
  record ascl:1610.016

\bibitem[{J. {Sanchez-Bermudez} {et~al.}(2022){Sanchez-Bermudez}, {Alberdi},
  {Sch{\"o}del}, \& {Sivaramakrishnan}}]{Sanchez-Bermudez_2022}
{Sanchez-Bermudez}, J., {Alberdi}, A., {Sch{\"o}del}, R., \&
  {Sivaramakrishnan}, A. 2022, \bibinfo{title}{{CASSINI-AUTOMAP: a novel image
  reconstruction algorithm for infrared interferometry},} in Society of
  Photo-Optical Instrumentation Engineers (SPIE) Conference Series, Vol. 12183,
  Optical and Infrared Interferometry and Imaging VIII, ed. A.~{M{\'e}rand},
  S.~{Sallum}, \& J.~{Sanchez-Bermudez}, 121831K, \dodoi{10.1117/12.2629488}

\bibitem[{A. {Sivaramakrishnan} {et~al.}(2023){Sivaramakrishnan}, {Tuthill},
  {Lloyd}, {Greenbaum}, {Thatte}, {Cooper}, {Vandal}, {Kammerer},
  {Sanchez-Bermudez}, {Pope}, {Blakely}, {Albert}, {Cook}, {Johnstone},
  {Martel}, {Volk}, {Soulain}, {Artigau}, {Lafreni{\`e}re}, {Willott},
  {Parmentier}, {Ford}, {McKernan}, {Vila}, {Rowlands}, {Doyon}, {Beaulieu},
  {Desdoigts}, {Fullerton}, {De Furio}, {Goudfrooij}, {Holfeltz}, {LaMassa},
  {Maszkiewicz}, {Meyer}, {Perrin}, {Pueyo}, {Sahlmann}, {Sohn}, {Teixeira}, \&
  {Zheng}}]{AMI2023}
{Sivaramakrishnan}, A., {Tuthill}, P., {Lloyd}, J.~P., {et~al.} 2023,
  \bibinfo{title}{{The Near Infrared Imager and Slitless Spectrograph for the
  James Webb Space Telescope. IV. Aperture Masking Interferometry},} \pasp,
  135, 015003, \dodoi{10.1088/1538-3873/acaebd}

\bibitem[{M. {Stalevski} {et~al.}(2017){Stalevski}, {Asmus}, \&
  {Tristram}}]{Stalevski2017}
{Stalevski}, M., {Asmus}, D., \& {Tristram}, K. R.~W. 2017,
  \bibinfo{title}{{Dissecting the active galactic nucleus in Circinus - I.
  Peculiar mid-IR morphology explained by a dusty hollow cone},} \mnras, 472,
  3854, \dodoi{10.1093/mnras/stx2227}

\bibitem[{M. {Stalevski} {et~al.}(2019){Stalevski}, {Tristram}, \&
  {Asmus}}]{Stalevski2019}
{Stalevski}, M., {Tristram}, K. R.~W., \& {Asmus}, D. 2019,
  \bibinfo{title}{{Dissecting the active galactic nucleus in Circinus - II. A
  thin dusty disc and a polar outflow on parsec scales},} \mnras, 484, 3334,
  \dodoi{10.1093/mnras/stz220}

\bibitem[{K.~R.~W. {Tristram} {et~al.}(2007){Tristram}, {Meisenheimer},
  {Jaffe}, {Schartmann}, {Rix}, {Leinert}, {Morel}, {Wittkowski},
  {R{\"o}ttgering}, {Perrin}, {Lopez}, {Raban}, {Cotton}, {Graser}, {Paresce},
  \& {Henning}}]{Tristram2007}
{Tristram}, K.~R.~W., {Meisenheimer}, K., {Jaffe}, W., {et~al.} 2007,
  \bibinfo{title}{{Resolving the complex structure of the dust torus in the
  active nucleus of the Circinus galaxy},} \aap, 474, 837,
  \dodoi{10.1051/0004-6361:20078369}

\bibitem[{K.~R.~W. {Tristram} {et~al.}(2022){Tristram}, {Impellizzeri},
  {Zhang}, {Villard}, {Henkel}, {Viti}, {Burtscher}, {Combes},
  {Garc{\'\i}a-Burillo}, {Mart{\'\i}n}, {Meisenheimer}, \& {van der
  Werf}}]{Tristram2022}
{Tristram}, K. R.~W., {Impellizzeri}, C.~M.~V., {Zhang}, Z.-Y., {et~al.} 2022,
  \bibinfo{title}{{ALMA imaging of the cold molecular and dusty disk in the
  type 2 active nucleus of the Circinus galaxy},} \aap, 664, A142,
  \dodoi{10.1051/0004-6361/202243535}

\bibitem[{R.~B. {Tully} {et~al.}(2009){Tully}, {Rizzi}, {Shaya}, {Courtois},
  {Makarov}, \& {Jacobs}}]{Tully2009}
{Tully}, R.~B., {Rizzi}, L., {Shaya}, E.~J., {et~al.} 2009,
  \bibinfo{title}{{The Extragalactic Distance Database},} \aj, 138, 323,
  \dodoi{10.1088/0004-6256/138/2/323}

\end{thebibliography}
\bibliographystyle{aasjournalv7}



\end{document}